%% file: main.tex
\documentclass[twocolumn]{aastex63}

\usepackage{booktabs, multirow} % for borders and merged ranges
\usepackage{xcolor,colortbl}
\usepackage{changepage,threeparttable} % for wide tables
\usepackage{verbatim} 
\usepackage{gensymb}
\usepackage{svg}
\usepackage{shortbold}
\usepackage{xspace}
\usepackage{natbib}
\usepackage{amsmath}

\DeclareMathOperator*{\argmin}{arg\,min}

\renewcommand{\alphaB}{\alpha B}

\newcommand{\hinode}{{\it Hinode}/SOT-SP\xspace}
\newcommand{\hmi}{{{\it SDO}/HMI}\xspace}
\newcommand{\hinp}{HinodeP\xspace}
\newcommand{\hmip}{HMIP\xspace}
\newcommand{\hmis}{\hmi Stokes\xspace}
\newcommand{\synthm}{SynthIA\xspace}
\newcommand{\synthp}{SynodeP\xspace}

\newcommand{\aia}{{{\it SDO}/AIA}\xspace}
\newcommand{\gauss}{Mx~cm$^{-2}$\xspace}

\definecolor{darkgreen}{HTML}{008800}

\submitjournal{ApJ}

\shorttitle{\synthm: A Synthetic H Inversion Approximation}
\shortauthors{Higgins et al.}

\graphicspath{{./}}

\begin{document}

\title{\synthm: A Synthetic Inversion Approximation for the Stokes Vector Fusing SDO and Hinode into a Virtual Observatory}

\correspondingauthor{Richard Higgins}
\email{relh@umich.edu}

\author[0000-0002-6227-0773]{Richard E. L. Higgins}
\affiliation{University of Michigan, Department of Electrical Engineering and Computer Science,
Ann Arbor, MI}

\author[0000-0001-5028-5161]{David F. Fouhey}
\affiliation{University of Michigan, Department of Electrical Engineering and Computer Science,
Ann Arbor, MI}

\author[0000-0003-0176-4312]{Spiro K. Antiochos}
\affiliation{NASA GSFC,
Silver Spring, MD}

\author[0000-0003-3571-8728]{Graham Barnes}
\affiliation{NorthWest Research Associates,
Boulder, CO}

\author[0000-0003-2110-9753]{Mark C.M. Cheung}
\affiliation{Lockheed Martin Solar and Astrophysics Laboratory, Palo Alto, CA}

\author[0000-0001-9130-7312]{J. Todd Hoeksema}
\affiliation{Stanford University,
Stanford, CA}

\author[0000-0003-0026-931X]{K. D. Leka}
\affiliation{NorthWest Research Associates,
Boulder, CO}

\author[0000-0002-0671-689X]{Yang Liu}
\affiliation{Stanford University,
Stanford, CA}

\author[0000-0003-1522-4632]{Peter W. Schuck}
\affiliation{NASA GSFC,
Silver Spring, MD}

\author[0000-0001-9360-4951]{Tamas I. Gombosi}
\affiliation{University of Michigan,
Department of Climate and Space,
Center for Space Environment Modelling,
Ann Arbor, MI}

\input{sec_abstract} \label{sec:abstract}


\section{Introduction}
\label{sec:introduction}
\input{sec_introduction}

\section{Methods}
\label{sec:methods}
\input{sec_methods} 

\section{Experiments}
\label{sec:experiments} 
\input{sec_experiments} 

\section{Results}
\label{sec:results}
\input{sec_results}

\section{Discussion and Conclusions}
\label{sec:conclusions}
\input{sec_conclusions} 

\acknowledgments
\input{sec_acknowledgements} \label{sec:acknowledgements}

\bibliography{main}{}
\bibliographystyle{aasjournal}



\end{document}

%% file: sec_abstract.tex
\begin{abstract}
Both NASA's Solar Dynamics Observatory (SDO) and the JAXA/NASA Hinode mission include spectropolarimetric instruments designed to measure the photospheric magnetic field. SDO's Helioseismic and Magnetic Imager (HMI) emphasizes full-disk high-cadence and good spatial resolution data acquisition while Hinode's Solar Optical Telescope Spectro-Polarimeter (SOT-SP) focuses on high spatial resolution and spectral sampling at the cost of a limited field of view and slower temporal cadence. This work introduces a deep-learning system named \synthm (Synthetic Inversion Approximation), that can enhance both missions by capturing the best of each instrument's characteristics. We use \synthm to produce a new magnetogram data product, \synthp (Synthetic Hinode Pipeline), that mimics magnetograms from the higher spectral resolution Hinode/SOT-SP pipeline, but is derived from full-disk, high-cadence, and lower spectral-resolution SDO/HMI Stokes observations. Results on held-out data show that SynodeP has good agreement with the Hinode/SOT-SP pipeline inversions, including magnetic fill fraction, which is not provided by the current SDO/HMI pipeline. SynodeP further shows a reduction in the magnitude of the 24-hour oscillations present in the SDO/HMI data. To demonstrate \synthm's generality, we show the use of SDO/AIA data and subsets of the HMI data as inputs, which enables trade-offs between fidelity to the Hinode/SOT-SP inversions, number of observations used, and temporal artifacts. We discuss possible generalizations of \synthm and its implications for space weather modeling. 
This work is part of the NASA Heliophysics DRIVE Science Center (SOLSTICE) at the University of Michigan under grant NASA 80NSSC20K0600E, and will be open-sourced.
\end{abstract}

%% file: sec_introduction.tex
The underlying source of energy for essentially all solar activity and space weather is the Sun's magnetic field. Since the magnetic field is responsible for high energy events, such as flares and coronal mass ejections, precisely measuring the photospheric and coronal magnetic field is of premier importance to understanding the physics behind space weather and improving space weather forecasting systems.  Since the magnetic field is presently difficult to measure routinely in the corona, where the energy for events is believed to be stored, models for space weather phenomena routinely invoke the photospheric magnetic field as a boundary condition.  Specifically, static models of the solar corona, from potential-field extrapolations to nonlinear force-free field models, rely upon estimates of the line-of-sight and full vector photospheric magnetic field respectively~\cite[cf.,][and references therein]{wiegelmann2021solar}.  Temporally-evolving models such as those based on magnetohydrodynamics with self-consistent turbulence (MHD)~\cite[cf.,][]{van2014alfven, Lionello2014}, magneto-frictional (MF) evolution and data-driving or data-assimilation approaches~\citep[cf.,][]{cheung2012method,hayashi2021coupling}, require precise and consistent estimates of this boundary, whether the model focuses on just the upper solar atmosphere or the full heliosphere \cite[cf.,][]{gombosi2018extended}. Beyond the models, empirical investigations into solar energetic phenomena rely on the photospheric field and its observed evolution for physical insights into the sources, causes, and effects of energetic events -- insights that then guide the modeling efforts themselves.

Given the importance of the photospheric magnetic field as boundary information, multiple facilities, both space and ground-based, have been constructed over the years to provide the needed estimates, trading off between spatial and spectral resolution and sampling, field of view, and cadence. Recording the polarized state of the light in magnetically sensitive spectral lines formed in the photosphere using the Stokes formalism, many instruments are coupled with a pipeline that processes the data to provide a standardized data product.  These pipelines often invoke the Milne-Eddington (ME) atmosphere \citep{Unno1956,Rachkovsky62} to best explain the observed data. ME-based inversions are a simplification that assume  a magnetic field and thermodynamic parameters that are constant with optical depth, and have been shown to generally provide appropriate averages of the magnetic vector as well as kinematic and thermodynamic parameters \citep{WestendorpPlaza_etal_1998,borrero2014comparison}, up to the inherent 180$^\circ$ ambiguity in the plane-of-sky azimuthal angle.

In order to ensure robustness for any study using these photospheric magnetic field estimates, they must be quantitatively precise over the whole photospheric boundary, and ideally in time as well as in space. These requirements place severe demands on the measurements. Yet as discussed, the magnetic fields are not observed directly, but are inferred through a complex inversion process.
The complexity of the inversion has led to well-known, long-standing inconsistencies between essentially all estimates of the photospheric field from different instruments, despite sophisticated and thorough instrument calibrations.  In spite of these efforts, producing full-Sun temporally-consistent quantitatively-defendable bias-free maps of the physical (heliographic) photospheric vector magnetic field remains one of the outstanding challenges in solar physics.

In this work we use machine learning (ML) to fuse two of these sensors and pipelines, creating a hybrid ``virtual'' observatory, specifically combining the {\it Helioseismic and Magnetic Imager} \citep[HMI;][]{Schou2012} on the {\it Solar Dynamics Observatory} \citep[SDO;][]{PesnellThompsonChamberlin2012} with the {\it Solar Optical Telescope-Spectro-Polarimeter} \citep[SOT-SP;][]{tsuneta2008solar} on the {\it Hinode} satellite \cite[]{kosugi2007hinode}. These instruments have complementary strengths. \hmi observes the full disk with $1.0\arcsec$ resolution (and $0.5\arcsec$ sampling) at a high cadence (e.g., $12$min for the {\tt hmi.S\_720s} data series we use). This field of view and cadence comes at the price of measuring the Stokes vector with somewhat coarse spectral sampling at only six wavelengths centered on one magnetically sensitive spectral line. \hinode, on the other hand, has far higher spectral resolution, and samples the Stokes vectors at 112 wavelengths across two spectral lines and additionally has a better spatial resolution of $0.3\arcsec$ per pixel; these resulting data are likely to more accurately represent the physical state of the photosphere. In exchange, however, \hinode's field of view is substantially smaller and employs a scanning-slit spectrograph. In order to obtain a spatial ``map'' not complicated by substantive evolution during the scan, often only the central portion of an active region is covered. 

\hmi and \hinode use different inversion pipelines to produce magnetograms. \hmi's pipeline uses a variant of VFISV~\citep{borrero2011vfisv}, described in \cite{Centeno2014}, to convert HMI Stokes observations to ME-based output. \hinode's pipeline performs the ME-based inversion using MERLIN~\citep{Lites2006}. It is important to note that one could use MERLIN on \hmi's data and vice-versa. To avoid conflating instruments (which produce Stokes spectropolarimetric data), inversion procedures (which convert Stokes data to magnetic field parameters), and pipelines (which combine instrument output, calibration modules, inversion techniques, and other steps to produce magnetic field parameters), we refer to: {\bf \hmis} as the Stokes filtergrams recorded by \hmi; {\bf \hmip} as the \hmi ME-inversion produced by the {\bf P}ipeline variant of VFISV on \hmis; and {\bf \hinp} as the ME-inversion produced by MERLIN on \hinode's Stokes filtergrams in the \hinode~{\bf P}ipeline. We will define the precise data series when we introduce the datasets used.

In this paper, we expand on our prior work, \citep[][hereafter H21]{higgins2021fast}. Our basic hypothesis is that information about the  magnetic field values obtained by \hinp is hidden but present in \hmis (potentially over multiple pixels), and that machine learning can learn to recover this information, given paired training samples. 

Our system, {\bf \synthm: the Synthetic Inversion Approximation}, is a multi-instrument, generally applicable framework for producing synthetic estimates of the output from the inversion of Stokes vectors from filtergrams and other instruments. In this paper, we propose \synthm and instantiate a version of it as a learnt mapping from \hmis to \hinp, yielding the new data product, {\bf \synthp: Synthetic Hinode {\bf P}ipeline}.
Once trained, \synthm enables emulations of \hinp on full-disk images at HMI cadence on the HMI grid and resolution. The resulting system offers the potential to enhance the value of both missions through the joint fusion of their data products. To focus our contribution, we limit ourselves to intrinsic field strength $B$, fill fraction $\alpha$ (and their product $\alphaB$), as well as plane of sky inclination $\gamma$ and azimuth $\psi$. However, our underlying technique has shown promise predicting kinematic and thermodynamic parameters of the photosphere.

Training \synthm requires a joint training set that aligns observations and inversion pipeline outputs (i.e., for \synthp, \hmis and \hmip). Due to inaccuracy in Hinode pointing information, we spatially align the data to a common (\hmi) grid with computer vision techniques \citep{lowe2004,Farneback03}.
For a given training set, our learned inversion approach is largely based on H21, which emulated the \hmi pipeline using a U-Net \citep{ronneberger2015unet} combined with regression-by-classification \citep[e.g.][]{Ladicky14b,Wang15}. This approach treats the problem as predicting a distribution over a set of discrete values, enabling uncertainty quantification and the integration of additional cues after training. \synthm  generalizes this approach across instruments.

We evaluate how well the system performs this mapping with multiple datasets. Our datasets are split chronologically (i.e., data used to fit the deep network parameters are separated by multiple months from the data used to test the network) to avoid data leakage \citep{Galvez2019,Liu2021}: many aspects of the Sun's evolution are slow and deep nets can memorize vast datasets. Performance on data near the training set cannot distinguish between generalization (approximating the right underlying function) and memorization (transferring from a slightly deformed copy of the same active region from a few days earlier). 

Our results show that \synthp has substantially better agreement with \hinp compared to with \hmip, both quantitatively as measured by mean absolute error (MAE) and qualitatively (at least to our eyes). Moreover, \synthp includes separated $\alpha$ and $B$, unlike \hmip, and shows good agreement with \hinp.
Our results on full-disk find broad agreement with \hmip but, like \cite{Hoeksema2014} and \cite{dalda2017statistical}, we also find systematic differences. We demonstrate the generality of \synthm by using a variety of inputs including data from the Atmospheric Imaging Assembly aboard SDO \citep[\aia;][]{lemen2011atmospheric}, leading to multiple \synthm variants with certain tradeoffs. We evaluate these variants in terms of both fidelity to \hinp as well as presence of temporal artifacts that are caused by SDO's 24-hour periodic orbit (captured by {\tt OBS\_VR}) as described by~\cite{Hoeksema2014}.

The \synthp outputs created by 
\synthm further offer qualitative benefits over both \hmip and \hinp. Compared to \hmip, \synthm determines both $\alpha$ and $B$ and has less strong (in magnitude) 24-hour oscillations, while also more closely resembling \hinp that is obtained from more detailed inputs. Compared to \hinp, \synthm can produce full-disk results continuously at the far higher cadence of \hmip, which enables monitoring multiple active regions simultaneously. Moreover, \synthm can operate with a variety of inputs, including \hmis only, \aia only, and \aia with subsets of the Stokes vector (or signals thereof). This paradigm enables obtaining estimates of field parameters even without a detailed physical model (e.g., UV/EUV images plus Stokes V and \hmi pseudo-continuum). We show that this approach can use UV/EUV data plus a single disambiguation bit from Stokes observations. These results may reduce data transmission requirements and suggests that future instruments may be able to ``add-on'' some magnetogram sensing at the cost of a few Stokes components.

From a science perspective, \synthm offers a way to empirically quantify the extent to which detailed information about the magnetic field is present in observations. Some results are counter-intuitive: \synthm's success at splitting $\alpha$ and $B$  (as is done in the \hinode pipeline) suggests that \hmip may not extract the full information from the spectra. Others, such as a bimodal splitting of inclination angle when conditioning on \aia data, agree with known physics, but serve as a data-driven verification.

% --- --- Pipeline --- ---
\begin{figure*}[t!]
\includegraphics[width=\linewidth]{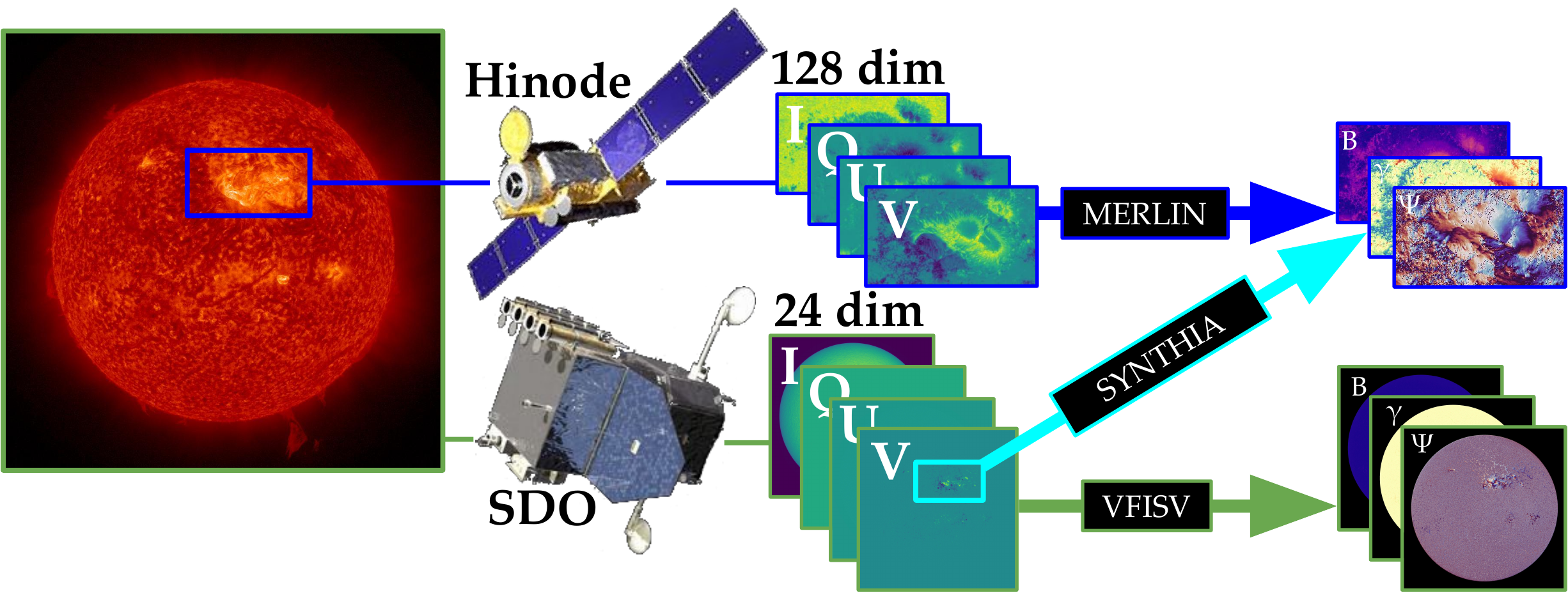}
\caption{\synthm is a model which inputs \hmis filtergrams (Section \ref{sec:input}) and learns to produce Hinode/SOT-SP Pipeline Inversions (\hinp) (Section \ref{sec:targets}), using a U-Net (Section \ref{sec:architecture_and_training}). \synthm yields a data product that we name \synthp, which are full-disk \hinode-like synthetic inversions on an \hmi grid that can be obtained with \hmi observables. We diagram a structure for how complementary sensors from different satellites can be combined to create new synthetic data products that may be useful for myriad downstream modelling tasks.}
\label{fig:pipeline}
\end{figure*}
% --- --- Pipeline --- ---

%% file: sec_methods.tex
The primary focus of this work is the training of a cross-satellite prediction method that maps \hmis observations to \hinode MERLIN Milne-Eddington inversion outputs originally produced using the polarized light recorded by \hinode (Figure \ref{fig:pipeline}). Our method, \synthm, is comprised of multiple convolutional neural networks (CNN) that each estimate a single parameter of the photospheric magnetic field. \synthm is based off of H21, which similarly uses a U-Net \cite[]{ronneberger2015unet} to produce maps of inversion targets from inputs within the {\it same} inversion pipeline, \hmi. Here, we use \synthm to generate a new data product, \synthp, that combines aspects of both \hmi and \hinode. 

In this section, we review \synthm generally, especially with an eye to the settings used to produce \synthp as a specific data product. To ground the description, we first discuss the inputs (Section~\ref{sec:input}) and outputs (Section~\ref{sec:output}) used in \synthm models presented here. 
We then discuss the components of building \synthm. Given data, the two required components are a neural net architecture (Section~\ref{sec:architecture_and_training}) and training optimization objective (Section~\ref{sec:objective}). However, unless the data are pre-aligned (e.g., as \hmis and \hmip naturally are),  \synthm requires the data to be registered (Section \ref{sec:aligning_hmi_and_hinodesp}). We conclude with computational implementation details about the neural net models (Section \ref{sec:implementation_details}).

\subsection{Input}
\label{sec:input} 

As a cross-satellite prediction method, we explore a variety of inputs. Our \synthp data product is based on \hmis observations. We  represent this as a 25-channel image consisting of:
24 channels of the four components of the Stokes vector ($I, Q, U, V$) observed across 6 bandpasses (the {\tt hmi.S\textunderscore720s} series); and 1 channel of pseudo-continuum image (the {\tt hmi.Ic\textunderscore720s} series). H21 also used 3 metadata channels, including heliographic coordinates and a flag labeling on-disk pixels. We found that this can lead to artifacts on full disk outputs (due to a dearth of Hinode polar scans in our training set), and so we omit metadata.

Since the method is learning-based, it is agnostic to the number of inputs so long as they can be aligned to a common grid. We thus experiment with other inputs, including multiple subsets of observations from \hmi and \aia.
Our \aia data consists of nine observations from  extreme ultraviolet (94, 131, 171, 194, 211, 304, 335\AA) and ultraviolet (1600, 1700\AA) channels from the {\tt aia.lev1{\textunderscore}euv{\textunderscore}12s} series. The channels are concatenated, and so different \synthm models require different input sizes 
(e.g., IQUV, continuum, and \aia UV/EUV data would have 34 channels). 

\subsection{Output}
\label{sec:output} 

Irrespective of input, the outputs of \synthm models throughout the paper represent \hinp parameters.
As in H21, rather than tune multiple losses, we estimate each of the paired inversion parameters separately, with individual networks. 
Since the angular resolution of the \hmi grid ($0.5$\arcsec) and \hinode grid (${\approx} 0.3$\arcsec) differ, resampling is unavoidable. We choose to work with the coarser resolution \hmi grid, since we aim to provide a method that can aid \hmi inversions by emulating a more spectrally capable system. We see super-resolution to \hinode's higher spatial resolution as orthogonal to our contributions and leave this to future work.

\subsection{Architecture and Training}
\label{sec:architecture_and_training} 

Inputs are passed through a UNet, mapping them to a 1-channel same-sized image output, where each pixel is a value predicted by the network for that corresponding location in the input. Following H21, the first half of the network is an encoder which extracts features at decreasing spatial resolution but increasing spatial extent. The second half is then a decoder which combines these extracted features from low resolution to high, resulting in an output that matches the spatial size of the initial input. The two are connected throughout via skip connections that  permit the re-use of high-resolution encoding information during decoding. Though both the MERLIN and VFISV inversions do not consider spatial extent, we still include it to help the network to discriminate between ambiguous inputs (e.g., distinguishing between the centers of sunspots and quiet regions) and to facilitate the use of information spread over multiple pixels. To reduce artifacts encountered between predicted tiles in H21, we change the convolutional filters within encoder and decoder blocks to include reflective rather than zero padding. 

Identically to H21, we train each model by minimizing a per-pixel loss (Equation \ref{eqn:objective}) with respect to $\thetaB$ via stochastic gradient descent \cite[]{robbins1951stochastic}, using the AdamW optimizer \cite[]{loshchilov2017decoupled}, with learning rate $10^{-4}$, $\epsilon=10^{-4}$, and weight decay $3\times10^{-7}$. We similarly scheduled optimization by observing results on a held-out validation set. During optimization, we monitor the number of consecutive epochs that occur without validation loss reduction. After four, we halve the learning rate; after six, we terminate training. This differs from H21 due to the decreased size of the datasets and was determined empirically by observing results on a held-out validation set. 

Our network has over 10 million trainable parameters and our code and models will be made open source. We trained our system using the PyTorch \cite[]{paszke2019pytorch} neural network framework.

\subsection{Objective}
\label{sec:objective}

We experiment with different training objectives, and compare them against one another in Section \ref{sec:experiments}. All settings train with $N$ pairs of inputs and associated/aligned targets $\{\XB^{(i)},\YB^{(i)}\}$, with height $H$ and width $W$. All settings also similarly attempt to find a selection of $\thetaB$ (a variable representing all trainable network parameters) that minimizes an aggregated per-pixel loss function shown here:   

\begin{equation}
\label{eqn:objective}
\argmin_{\thetaB}
\sum_{i=1}^N \sum_{p} 
 \mathcal{L}(f(\XB^{(i)};\thetaB)_{p},\YB^{(i)}_{p})
\end{equation}

\vspace{2mm}
Classification is the first objective we consider for mapping 25 input channels to 1 output channel at each pixel. Following H21, we consider $f$ that maps $\mathbb{R}^{H \times W \times 25} \to \mathbb{R}^{H \times W \times 80}$. Here the 80-dimensional output is a series of bins that correspond to linearly subdividing the possible target values. The target is created with probability mass allocated across two adjacent bins such that the expectation of the bins is the real value, plus label smoothing, as done in~\cite{Ladicky14b,higgins2021fast}. This method is trained with the Kullback Leibler (KL) divergence \cite[]{kullback1951information} between the predicted distribution and target distribution
and is identical in output formulation to H21. The training objective is per-pixel, in that no terms tie together pixels when producing outputs. Although the network jointly produces outputs for the entire input height by width, no loss penalizes any joint predictions, and every pixel only depends on local windows of input. At inference time, one only needs to provide the network an input patch of at least $32 \times 32$ pixels, and each pixel's output is a separate expectation at the most likely bin in the per-pixel probability distribution.

An alternate objective that can be used is Regression. 
In this setting, $f$ is a function mapping $\mathbb{R}^{H \times W \times 25} \to \mathbb{R}^{H \times W \times 1}$. Below, we set $\hat{\yB}=f(\XB^{(i)};\thetaB)_p$ and target $p = \YB^{(i)}_p$, casting our problem as regressing a target-value $\vB \in \mathbb{R}$. We then penalize estimates $\hat{\yB}$ that deviate from the target value. One error is the Mean Squared Error (MSE), which measures the squared difference between $\hat{\yB}$ and $\pB$, or at the jth pixel $(\hat{\yB}_j - \pB_j)^2$.
The MSE thus penalizes predictions based on the square of differences across the image.

\subsection{Aligning HMI and SP Data}
\label{sec:aligning_hmi_and_hinodesp}

\synthm requires data to be aligned spatially. Thus, to produce the \synthp data product, we need to align \hmi and \hinode data. Since \hinode pointing information is not reliable enough to yield pixel-level correspondence, we obtain an alignment by fitting a transformation using computer vision techniques. 
The first component is a global transformation that localizes the \hinp cutout on the disk. The global transformation is followed by a per-pixel deformation that handles unmodeled phenomena, like evolution during \hinode's scans and slight rotation (since HMI's grid is ${\approx}\,0.1^\circ$ away from Solar north).

Our global transformation is fit on correspondences between \hmi and \hinode data that are extracted using SIFT~\citep{lowe2004}. SIFT finds features at distinctive parts of images at multiple scales, and describes them with gradient information near each distinctive region. Given two images, SIFT matching generates candidate correspondences. Unfortunately, since the matches are based on local windows, these correspondences are likely outlier-contaminated. The standard solution to SIFT outliers (e.g., see~\citep{Brown03}) is RANSAC~\citep{Fischler81}, which finds a set of inliers (i.e., a set of points that fit a model well) in the data and then fits a model via least-squares to those inliers. Point correspondences have the advantage of being robust to some intrinsic misalignment (e.g., due to evolution during \hinode's scan) because the features are local. The misalignment is minimized in the least-squares. 

We fit a scaling and translation model on point correspondences extracted from multiple data available in both \hmi observations and the \hinode MERLIN Level 2 data, specifically Continuum information (MERLIN {\tt Original\_Continuum\_Intensity}, {\tt hmi.Ic\_720}) as well as Stokes V information (MERLIN {\tt StokesV\_Magnitude} and the 6 Stokes V observations from {\tt hmi.S\_720s}). Our scaling model includes the non-unity aspect ratio of pixels in \hinode (${\approx}1.07$) and our RANSAC procedure additionally rejects models that are incompatible with the theoretical scale ratio between the two instruments.

The second component of the transformation is a per-pixel flow that is fit by optical flow, specifically using the method of~\cite{Farneback03}. We fit the flow on the input data with the most inliers, initialized with the global transformation. The optical flow aims to align the two images as well as possible while preferring simpler deformations.

Once an alignment has been fit, we can warp an observation on \hinode's grid to HMI. Our decision to interpolate \hinode ensures that \hmi data is never interpolated. We align \hinode by mapping each \hmi grid point in the corresponding (fractional) \hinode grid point and interpolating with a 3rd order spline fit on the \hinode data. Special care is needed for circular and fill-factor targets; this is discussed in Section~\ref{sec:targets}.

Not all \hinode observations can be aligned, and our process rejects many observations using quality thresholds that are determined empirically. We reject \hinode observations where {\tt Mechanical\_Slit\_Position} does not increase in a regular ascending pattern. During the rigid fitting procedure, we reject the alignment if fewer than 5 inliers can be found. Finally, post-fitting, we independently z-score the aligned continuum images from \hinode and \hmi (i.e., so their mean is zero and standard deviation is one); pairs where 20\% of the pixels have a difference of over $1.5$ are rejected as being likely poorly aligned. 

Of the 4719 observations we obtained during the period 1 January 2012 -- 31 December 2014, 2920 observations passed all the alignment requirements. Common alignment failures include (in addition to observations with sudden slit jumps) polar regions and small scans containing only granulation. While the small scans with only granulation may prove impossible to align with current techniques, we suspect the polar regions could be aligned by an alternate method using the limb. We leave this to future work.

\subsection{Implementation Details}
\label{sec:implementation_details}

Training on \hinode cutouts involves processing input images in a variety of sizes. However, all cutouts are smaller than $600$ pixels in size (resulting from re-scaling the \hinode observations to the \hmi grid) and fit entirely into GPU memory. The variable sizes pose no problem for the network, as it is fully convolutional. Testing on HMI data poses data challenges. As in H21, GPU memory constraints mean we must divide full-disk $4096\times4096$ \hmi images into $1024\times 1024$ pixel tiles when running inference. Our full-disk quantitative evaluations are done with non-overlapping tiles (leading to $16$ tiles) covering the full disk. To reduce slight border artifacts, stitched full disk outputs (e.g., such as Figure~\ref{fig:fig_full} are created using overlapping tiles. These are spaced every $512$ pixels (leading to $49$ tiles) and the middles are cut out. Inference on a GeForce RTX 2080 Ti GPU with 4352 CUDA cores, takes on average 300 ms per tile, assuming data is pre-loaded into system RAM. On an 8 GPU server, the time to read data off of a single SSD is the limiting factor in the network's operation. In aggregate, considering time spent moving data from system memory to GPU memory and time spent running the neural network, computing  a full-disk via 49 tiles takes 14.7 seconds to process on a single GPU.

%% file: sec_experiments.tex
% --- --- Fig Full --- ---
\begin{figure*}[t]
    \makebox[\linewidth]{
        \includegraphics[width=\linewidth]{fig_full.pdf}
    }
    \caption{Full-disk \synthp outputs for field strength ($B$), fill factor ($\alpha$), inclination angle ($\gamma$), and azimuthal angle ($\psi$) for 2016 May 16 06:24:00 TAI. \synthm uses full-disk \hmis measurements to produce \synthp, \hinp-like full-disk inversion outputs on the \hmi grid. \synthm cross-calibrates multiple sensors on different satellites, trains on data from small paired co-observed cutouts, and produces a new data product, \synthp that combines good qualities of both \hinp and \hmip (e.g., \hinp's recovery of magnetic fill factor and \hmip's full disk coverage). High-quality, full-disk results like these suggest \synthp might be appropriate for use in space weather forecasting pipelines that would benefit an instrument combining characteristics of both \hinp and \hmip.}
    \label{fig:fig_full}
\end{figure*}

We evaluate our cross-satellite prediction technique via a number of experiments. Our goals are to:
(a) assess how well the resulting \synthp matches \hinp in comparison to \hmip; (b) examine full-disk behavior of \synthp; (c) evaluate alternate approaches for \synthm; (d) assess \synthm variants that use different inputs; and (e) assess the temporal variability of \synthm outputs. We now describe the data used, its preparation and characteristics, as well as metrics used throughout. 

\subsection{Targets}
\label{sec:targets}

We focus our attention on magnetic field parameters: field strength, fill factor, inclination, and azimuth. All are produced by \hinp, and all but fill factor and true field strength are available from \hmip. H21 also demonstrated predictions of the kinematic (LOS Doppler Velocity) and thermodynamic (e.g., source function constant $S_0$) parameters of the ME atmosphere model; we omit these in the interest of focusing instead on how varying input data affects performance of the magnetic field inferences, specifically.

Particular care is needed when analyzing field-strength quantities since \hinp and \hmip treat fill factor differently. \hinp fits two parameters to describe the field magnitude: a field strength, representing intrinsic field strength, as well as a fill factor, representing the fraction of light entering a pixel caused by a magnetized atmosphere. \hmip, on the other hand, fixes fill factor to unity~\citep{Centeno2014} in its implementation of VFISV so that the returned parameter is the area-averaged field strength or magnetic flux density. This approach optimizes performance in regions where magnetized plasma fills the pixel, such as in sunspots: when the fill factor is indeed one, \hmip field strength matches intrinsic field strength. On the other hand, in plage and other regions with fragmented sub-resolution magnetic structures, \hmip's field does not represent field strength, as the fixed unity fill-factor is not appropriate.
\begin{enumerate}
\item {\it Magnetic Field Strength} ($B$), predicted from $0$ to $5000$ Gauss. Most pixels have a low field strength on the disk, but the rarer large values are most important to predict correctly for space weather. \hinp reports this as {\tt Field\_Strength}. Since \hmip does not fit a fill factor, quantitative comparisons between \hinp and \hmip field are not meaningful.
The difference between field strength $B$ and area-averaged field strength $\alphaB$ is best seen in plage and other areas with meaningful polarization outside active regions.
    
\item {\it Fill Factor} ($\alpha$), predicted from $0$ to $1$, which \hinp
reports as {\tt Stray\_Light\_Fill\_Factor}. \hmip reports a constant value of unity.
    
\item {\it Magnetic Flux Density, or Field $\times$ Fill Factor, or Area-Averaged Field Strength} ($\alpha B$), which we predict from $0$ to $5000$ \gauss. Since \hmip fixes the fill factor to unity, {\tt field} from {\tt hmi.ME\_720s\_fd10} is best described as $\alpha B$. For \hinp, $\alpha B$ can be computed; we perform the calculation on the pipeline output and then interpolate. 

\item {\it Inclination} ($\gamma$), predicted from $0^\circ$ to $180^\circ$. 
\hinp reports this as {\tt Field\_Inclination}, and \hmip reports this
as {\tt inclination}. Inclination is within $0^\circ$ and $90^\circ$ when the line-of-sight component of the field points toward SDO, and $90^\circ$ and $180^\circ$ when it points away. In regions with low polarization signal, inclination tends toward $90^\circ$. I.e., field leans to plane of sky because of the noise characteristics of the measurement \citep{LaBonte2004,Pevtsov_etal_2021}.
    
\item {\it Azimuth} ($\psi$), predicted from $0^\circ$ to $180^\circ$. 
Azimuth is the angle of the magnetic field vector projected onto the plane of the sky as observed by the instrument.
\hinp reports Azimuth as {\tt Field\_Azimuth}, and \hmip reports it
as {\tt azimuth} measured in degrees counter-clockwise from {\it up} on the CCD. We are operating with pre-disambiguation data products, so the azimuth is inherently $180^\circ$ ambiguous, hence the inversion of a single pixel cannot distinguish, for example, between $10^\circ$ and $190^\circ$. 

We interpolate Azimuth on the unit circle to avoid angular preferences. Specifically, we map each angle $\theta \in [0^\circ, 180^\circ]$ to a point on the unit circle $\sin(2\theta),\cos(2\theta)$. These data are interpolated, reprojected to the unit circle, and then remapped back to angles. This method mitigates a preference for $90^\circ$ that can occur if ambiguous angles were directly interpolated (e.g., the average of $10^\circ$ [alternatively $190^\circ$] and $160^\circ$ [alternatively $-20^\circ$] should be $175^\circ$ [or $-5^\circ$] rather than $85^\circ$).

When the transverse component of the field is weak, either because the field lies along the line of sight or because the overall magnetic field strength (polarization signal) is weak, the azimuth angle becomes highly uncertain. In general the magnetic field vector in the photosphere tends toward radial and simple morphologies of solar fields expand away from concentrations of strong field in an azimuthally-symmetric formation. These solar effects result in plane-of-sky azimuthal-angle preferences that depend on radial field direction and disk position; however, it is worth noting that the most interesting areas on the Sun are exactly those where the general rules do not apply. Unlike inclination angle, there is no obvious bias on azimuth for \hmip and \hinp, and thus no preferred value seen in low polarization region for azimuth. 
Azimuth from ME data is roughly equally distributed from 0 to 180 degrees.
\end{enumerate}

\subsection{Evaluation}
\label{sec:evaluation}

We evaluate models' abilities to predict these targets following H21, which evaluated across full-solar disks in emulating the \hmi VFISV inversion. Specifically, they used two metrics to compare model results, $\hat{y}$, against scalar targets, $y$. The first is Mean Absolute Error (MAE), an average of the per-pixel metric $|y-\hat{y}|$ across the entire dataset. Due to this metric being susceptible to outliers as an average measure across pixels, they employ a second metric, a percentage count of well-estimated pixels. This is the average measure of pixels that are withing a given distance, $t$, of the target value: $|y - \hat{y}|<t$. For simplicity, we adopt the values chosen in H21 for $t$, despite targeting a different inversion output. We also use the same approach to pick a threshold for fill-factor, which is not inferred by \hmip. Finally, we calculate these values in a circular fashion for azimuth (i.e., the error is $\min(|y-\hat{y}|,180-|y-\hat{y}|)$.

\subsection{Data Preparation}
\label{sec:datasets}

We experiment with two pipeline outputs (\hmip and \hinp) and two input data sources (the \hmi Stokes parameters and continuum information and \aia level 1 data). All SDO and Hinode data are publicly available on JSOC and HAO respectively. All generated data will be made public and assigned DOIs. 

We align all data onto the \hmi observation grid to minimize interpolation by potential users. Any downstream users that would need to interpolate the results would already be interpolating; if the data were aligned to another grid (e.g., a grid aligned exactly to solar north rotated by {\tt CROTA\_2}), interpolation would be needed to compare with raw \hmi data products. For simplicity, we rotate the image without interpolation 180$^\circ$ to nearly align with heliographic north (but do not align to {\tt CROTA\_2}). 

All datasets contain \hmi observations from {\tt hmi.S\_720s} and {\tt hmi.Ic\_720s} and inversion results from {\tt hmi.ME\_720s\_fd10}. The paired training datasets also contain \hinode MERLIN inversions, which are co-aligned to the \hmi grid via the method described in Section~\ref{sec:aligning_hmi_and_hinodesp}.

To test generality, the datasets also contain \aia observations. We use level 1 data \citep{lemen2011atmospheric} which has been post-processed to: convert $\textrm{DNs}$ to flux by dividing by exposure time; correct for instrument degradation via AIAPy~\citep{Barnes2020}; and re-bin so \aia data has the same angular resolution as \hmi. Our procedure aligns the imaging sensors, but \aia EUV channels capture photons that primarily originate in the Sun's corona, substantially above the photosphere. Since the phenomena occur at different locations, there is intrinsic misalignment between image features in the EUV observations and the photospheric magnetic field (except at disk center). The misalignment is height-dependent and thus channel-dependent: 1600\AA~and 1700\AA~have the least intrinsic misalignment since they originate closest to the photosphere. Channels originating in the corona (e.g., 335\AA) have worse misalignment. 

% --- --- Synode vs. Hinode MERLIN Inversion --- ---
\begin{figure*}[t]
    %\vspace*{-2cm}
    \makebox[\linewidth]{
        \includegraphics[width=\linewidth]{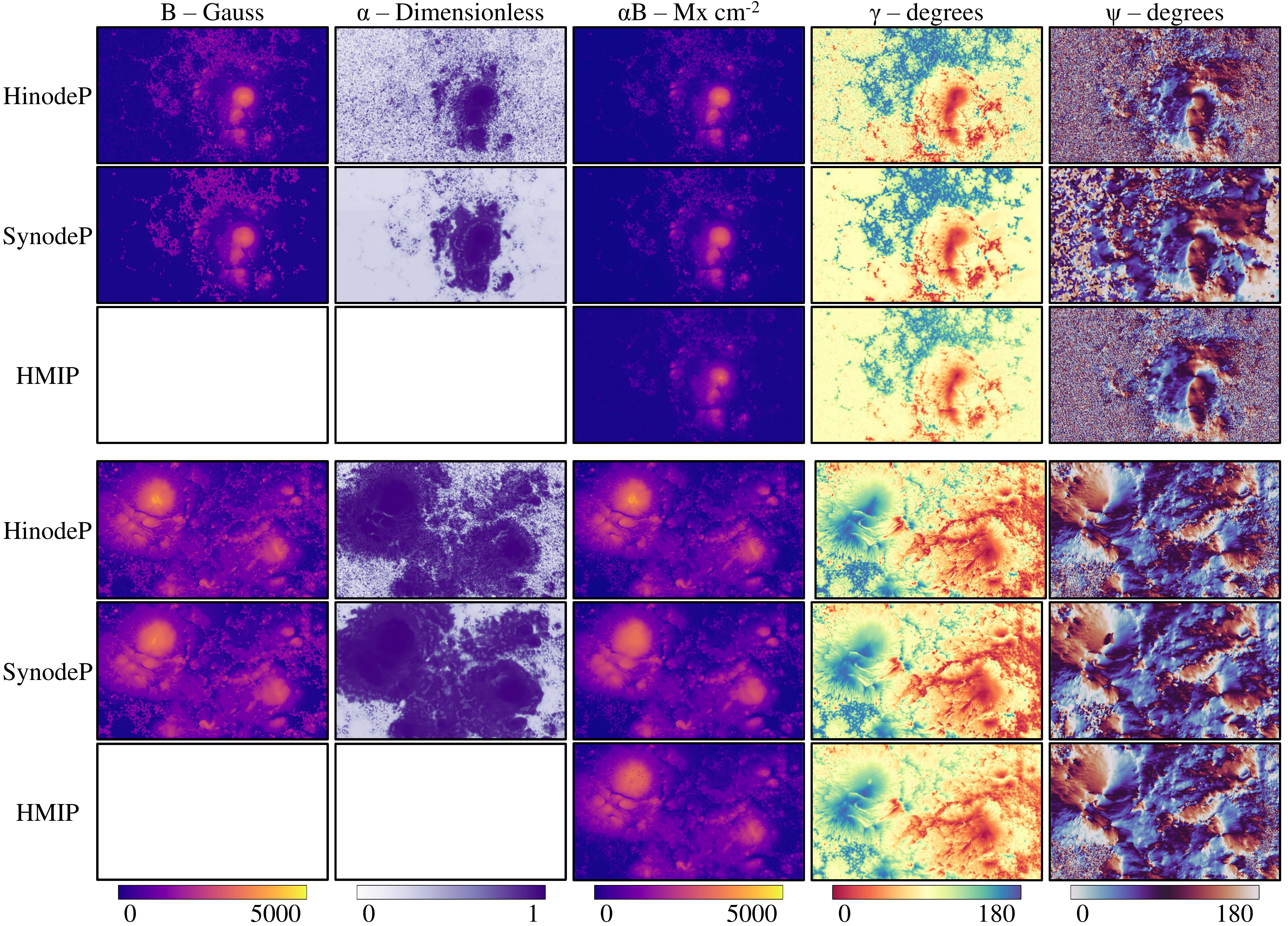}
    }
    \caption{Qualitative comparisons of \hinp inversions, \synthp synthetic inversions produced by \synthm, and \hmip inversions generated from two {\bf Paired SDO-Hinode} cutouts.
    Top: 2014 September 11 19:07:05 TAI; Bottom: 2014 October 24 08:45:05 TAI. Results for independent $\alpha$ and $B$ are not available for \hmip, since \hmip solves for their product. The comparison shows \synthp's ability to generate \hinp-like $B$ and $\alpha$ (first two columns), despite using \hmis data as input. The figure also shows \synthm models noisy azimuthal values qualitatively differently, as seen in \synthp in the last column.
    }
    \label{fig:hinode_synthm_vfisv}
\end{figure*}

\subsection{Datasets}

We work with three data sets for training and validating \synthm models. 

    {\it Paired SDO-Hinode (2012--2014)} is used for training \synthm models, and consists of cutouts of recorded {\it SDO} data that spatially and temporally match the 2920 \hinode observations from January 1, 2012 to December 31, 2012 that can be aligned by the method described in Section~\ref{sec:aligning_hmi_and_hinodesp}. We pick the corresponding \hmi observation based on the nominal \hinode scan start time. The data is split by timestamp into training (for fitting network parameters $\thetaB$), validation (for monitoring training progress and making design decisions but {\bf not} fitting $\thetaB$), and held-out testing (for evaluating performance). We use a 80/10/10\% train/validation/testing split. The training data ends 2014 June 04 and testing data begins 2014 August 11, with validation in the middle.
    
    {\it HMI 2016-All} is from H21, and was made from HMI full-disk data every other day starting January 1, 2016 - 31 December 2016. Hours and minutes were sampled randomly (from available options) and ignoring failures. In this work, we entirely use this dataset for testing how training on the above paired dataset translates to predicting with full-disk \hmi inputs.
    
    {\it 2016 Feb Sequence} is a subset of the Month-long sequence used in H21 and is obtained every hour for a two week period at the start of February 2016. We use this to evaluate the presence of temporal oscillations.

Our dataset selection and experiments are deliberately constructed to minimize issues with potential leakage due to the slow evolution of the Sun as discussed in~\cite{Galvez2019,Liu2021}. The paired SDO test data is at least 68 days after the final training sample (${\approx}2.5$ Carrington rotations) and the full-disk outputs we show are generated more than a year after the training data. We periodically show qualitative results on various other \hmi data; these results are from 2016 (and thus ${\ge}1.5$ years after the training data).

% --- --- End Synode vs. Hinode MERLIN Inversion --- ---
\begin{figure*}[t]
\centering
\includegraphics[width=\linewidth]{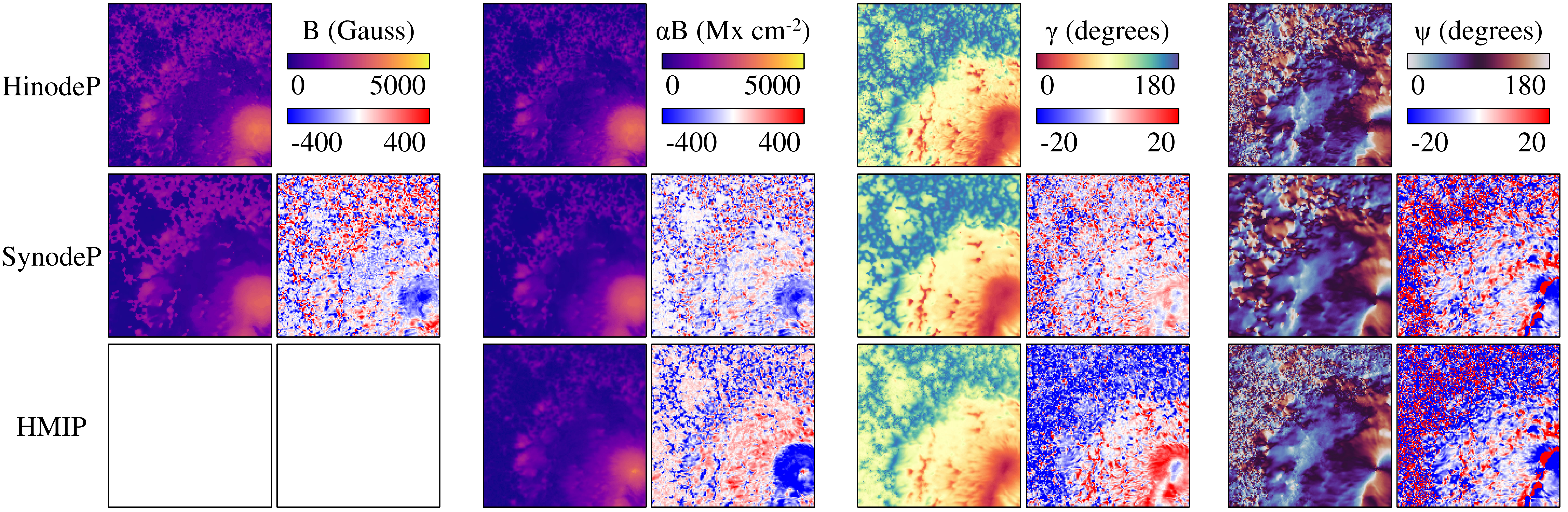}
\caption{Zoom-in on a 75\arcsec~region from Fig.~\ref{fig:hinode_synthm_vfisv}, 2014 September 11 19:07:05 TAI. \hinp, \synthp, and \hmip outputs are shown (except for $\alpha$, which is constant for \hmi). Difference maps visualize \synthp and \hmip as predictors of \hinp: blue/red are under/over-prediction. Note: \hmip does not produce $B$. Results for $\alphaB$, inclination angle, and azimuthal angle qualitatively show \synthp is closer to \hinp than \hmip is, as seen through less blue/red in difference maps.}
\label{fig:zoomin}
\end{figure*}

%% file: sec_results.tex
We now report results to evaluate the proposed \synthm system, focusing on both the \synthp data product it enables as well as variations that use different inputs and temporal analyses.

In Section \ref{sec:synode_vs_hinode}, we first test the performance of \synthp in predicting \hinp cutouts from the \textbf{Paired SDO-Hinode} dataset, using \hmip as a baseline. We then 
test how \synthp compares against both \hmip and \hinp on matching cutouts.

Having shown results on \hinp cutouts, we next evaluate \synthp full-disk outputs against our only other full-disk inversion output, \hmip, in Section \ref{sec:synode_vs_hmi}. For a better understanding of how our \synthp outputs differ, we further divide the full-disk inversion outputs from \hmip into pixel populations such as active regions and plage, and examine these specific regions in comparison to \synthp.

We next examine our design decisions in Section \ref{sec:experiments_ablations}, and explore a variety of modifications to \synthp, including regression as a target paradigm, the use of meta-information, and the prediction of field and fill factor independently versus together.

We then test the generality of our approach in Section \ref{sec:aia_for_prediction}, by exploring the introduction of \aia data. While still predicting \hinp as the desired output, we examine a few different options as input instead of \hmis~ -- using AIA only, adding AIA data, and combining AIA with polarization subsets of original \hmis data (e.g., just \textit{V} or just \textit{I \& V}).

We conclude by examining temporal trends in Section \ref{sec:network_behavior_across_time}.
\hmip is known to have temporal oscillations associated with the 24-hour orbit, and past work on learned emulation of the pipeline has reproduced these oscillations. By exploring how \synthp-produced outputs function across time, we attempt to isolate and explain what components might be the cause of oscillations.

\begin{deluxetable*}{lc@{~~~}ccccccc}[th]
\caption{Results on \textbf{Paired SDO-Hinode}, testing \synthp and \hmip against \hinp as ground truth. We evaluate with Mean Absolute Error (MAE) and percent of pixels within $t$ for each test set Hinode cutout. Values for $t$ are target specific, generated by scaling according to relative variances. For all targets, \synthp is closer to \hinp than \hmip is to \hinp.} 
\label{tab:hinode_results}
\tablewidth{0pt}
\scriptsize
\tablehead{\colhead{Target}&\colhead{Range}&\multicolumn2c{MAE}&\colhead{$t$}&\multicolumn2c{\% Within $t$}\\
&&\colhead{\synthp} &\colhead{\hmip} &\colhead{}&\colhead{\synthp}&\colhead{\hmip}}
\startdata
Field Strength ($B$) & [0,$5\!\times\!10^3$] Gauss & \bf 137.1 & 197.7 & 47 & \bf 39.7\% & 31.3\% \\
Fill Factor ($\alpha$) &[0,1] &\bf 0.11 &0.49 &0.1& \bf 54.9\% & 9.4\% \\
Fill Factor $\times$ Field Strength ($\alpha B$) &[0,$5\!\times\!10^3$] \gauss &\bf 72.4 &109.7 &47& \bf 60.5\% & 28.2\%\\
Inclination ($\gamma$) &[0,180] \degree &\bf 9.4 &12.5 &5& \bf 40.1\% & 31.1\%\\
Azimuth ($\psi$) &[0,180] \degree &\bf 24.0 & 28.3 &7& \bf 33.4\% & 26.3\%
\enddata
\end{deluxetable*}

% --- --- Synode / VFISV vs. Hinode MERLIN Comparison --- ---
\begin{figure*}[t]
\centering
\includegraphics[width=\linewidth]{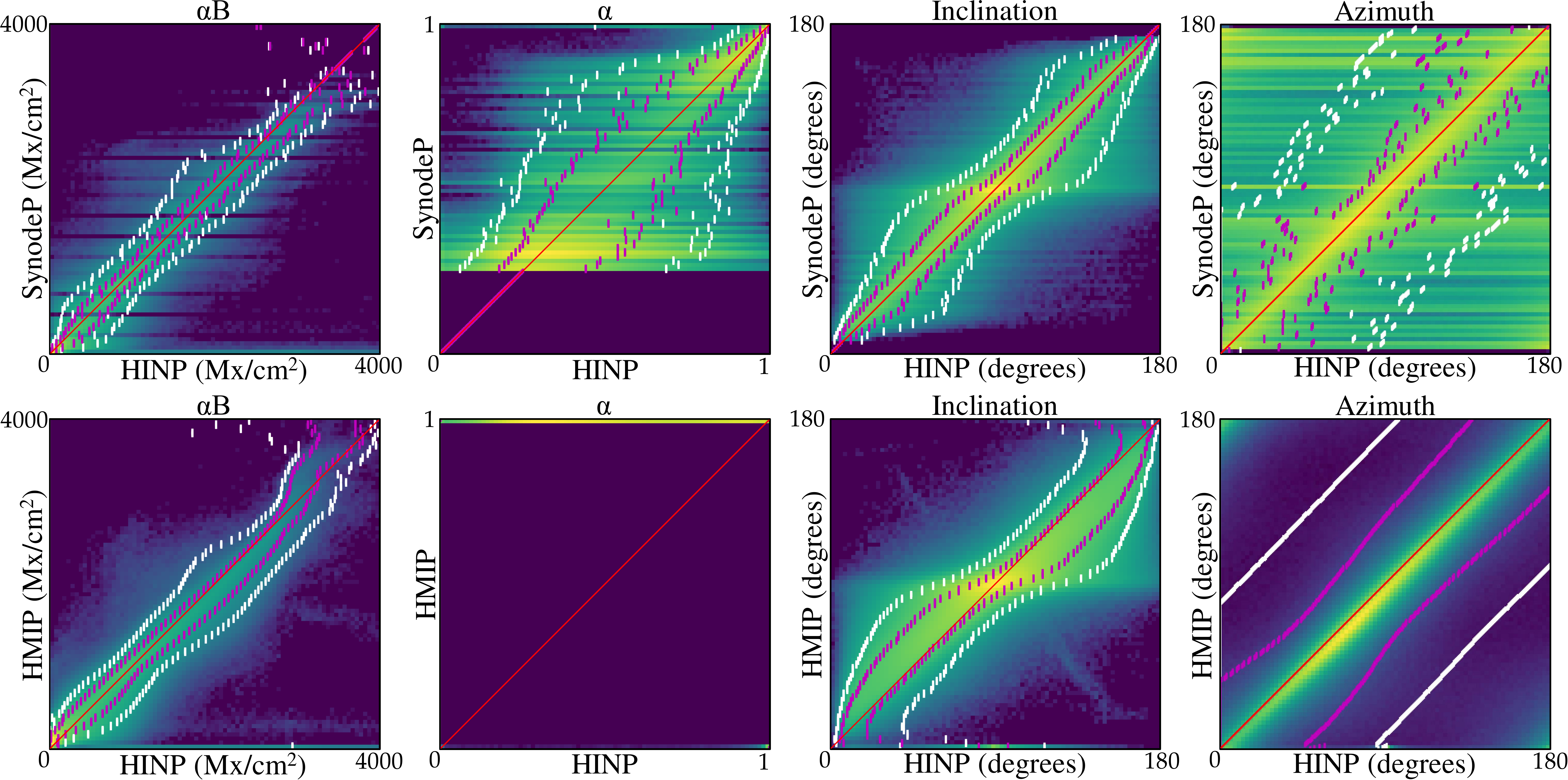}
\caption{Visualizing performance on \textbf{Paired SDO-Hinode}, comparing \synthp and \hmip against \hinp with {\bf log count} bi-variate histograms.  The line $y=x$ depicts perfect predictions, while pink/white dashes show bounds on the left and right (lower and upper) for empirical 67\%/95\% prediction CIs respectively. }
\label{fig:pairedSDO_histograms}
\end{figure*}

\subsection{Predicting \hinode MERLIN Inversions from \hmis data} \label{sec:synode_vs_hinode}

We begin by testing how well \synthp can reproduce \hinp using \hmis inputs. Our criterion for success is that \synthp should produce results that are closer to \hinp compared to \hmip output. 

We first show qualitative results on held-back data from the {\bf Paired SDO-Hinode} dataset in Figure~\ref{fig:hinode_synthm_vfisv} for the five targets. In all cases, there is generally good agreement, although results for all targets are slightly blurry (due to the imperfect alignment used for training). We report quantitative results on the {\bf Paired SDO-Hinode} dataset for both \synthp and \hmip in Table \ref{tab:hinode_results}, where \synthp shows substantially better agreement to \hinp than \hmip. These are best viewed with the bivariate log histograms of Figure~\ref{fig:pairedSDO_histograms}.

\synthp predictions of {\bf Field Strength/$B$} closely match \hinp. Since \hmip does not contain true field strength, we do not show a comparison. If one treats $\alphaB$ as $B$ there is generally good agreement in strongly polarized regions since these regions have $\alpha$ near unity. However, clear differences are visible in plage, as also reported by \cite{dalda2017statistical}. 
Despite using the same inputs as \hmip, \synthp is able to estimate the field reported by \hinp well. This is born out in quantitative results.

This trend continues with {\bf Fill Factor}/$\alpha$, although many islets outside of active regions are predicted as uniform. Given that $\alpha$ is not inferred by \hmip, the results shown could potentially be incorporated into the pipeline.
A prominent failure mode is \synthp's tendency to smooth out small regions with significant signal (small few-pixel islets of regions with high fill amidst a sea of low-fill pixels), as can be seen in both inclination and $\alpha$. The failure of \synthp to predict the islets may be driven by misalignment of these small regions at training time and is further discussed in Section~\ref{sec:limitations}. If these regions are frequently misaligned, the ground-truth may frequently be $90^\circ$ for inclination or close to zero for $\alpha$. Finally, \hinp contains $\alpha$ values less than 0.2, yet \synthp predictions are never that low, as seen in Figure \ref{fig:pairedSDO_histograms} (row 1, column 2). We hypothesize this is due to a combination of $\alpha$ being interpolated, reducing the likelihood of pure $\alpha=0$ pixels, as well as the aforementioned oversmoothing.
The quantitative results agree with the qualitative assessment, with a MAE of 0.12, covering ${\sim}10\%$ of the range. This is smaller than \hmip, even though the cutouts are focused on active region areas (where the \hmip $\alpha{=}1$ assumption is most likely to be valid).

These first two targets of {\bf Field Strength/$B$} and {\bf Fill Factor}/$\alpha$ are not produced by \hmip, so while \synthp's good agreement with \hinp is interesting, its ability to do better than \hmip is not surprising. We next examine three quantities that all systems produce, where \synthp continues to produce better agreement than \hmip. However, these generally require closer visual inspection, and we encourage looking at Figure~\ref{fig:zoomin}.

The first target produced by all systems is {\it Fill Factor $\times$ Field Strength} or $\alphaB$, where \synthp's predictions closely match the \hinp output. The results are also closer, quantitatively speaking, than \hmip. Just as in the case with intrinsic field strength, differences continue to be seen in plage, where \synthp and \hinp infer higher values compared to \hmip. In quiet regions, \hmip produces higher $\alphaB$ values than \hinp, while \synthp produces consistent values. This illustrates that \synthp is reproducing nuanced instrument- and pipeline-specific behavior. 

\begin{figure*}[t]
    %\vspace*{-2cm}
    \makebox[\linewidth]{
        \includegraphics[width=\linewidth]{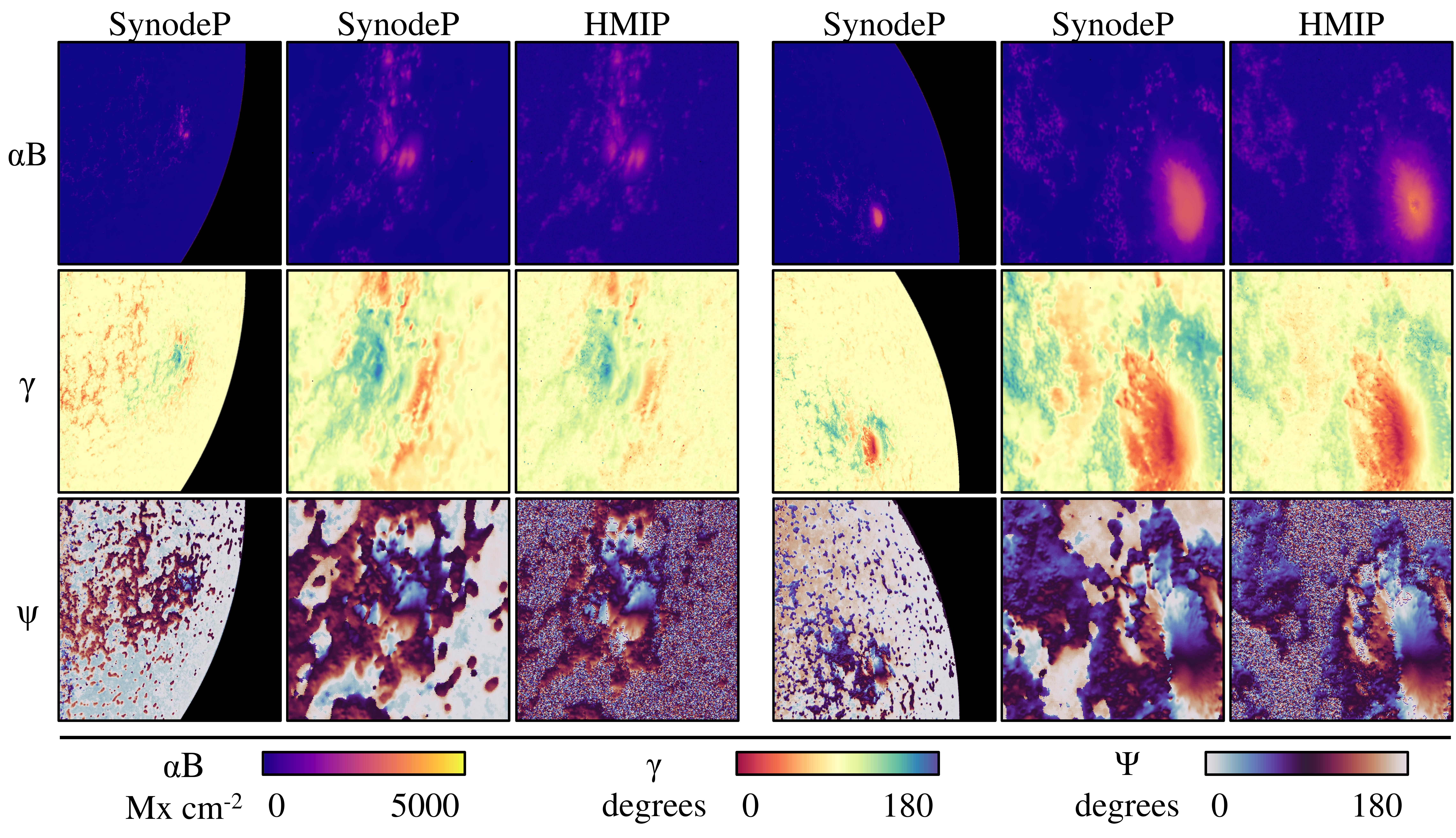}
    }
    \caption{\synthp compared to \hmip (left: 2016 May 14 16:00:00 TAI; right: 2016 May 24 09:36:00) on a disk containing a large active region. We show 512\arcsec and then 125\arcsec cutouts for \synthp on both days. \synthp shows remarkable agreement to \hmip, despite being trained against \hinp as a target. Besides azimuthal predictions in weak polarization areas (last row), qualitative differences are small. \synthp under-estimates $\alpha B$ compared to \hmip (first row, final two columns). The fact that \synthp closely models \hmip outside of \textbf{Paired SDO-Hinode} cutouts demonstrates its potential applicability as a full-disk synthetic inversion.}
    \label{fig:qualitative_hmi}
\end{figure*}

\begin{figure*}[t]
\centering
\includegraphics[width=\linewidth]{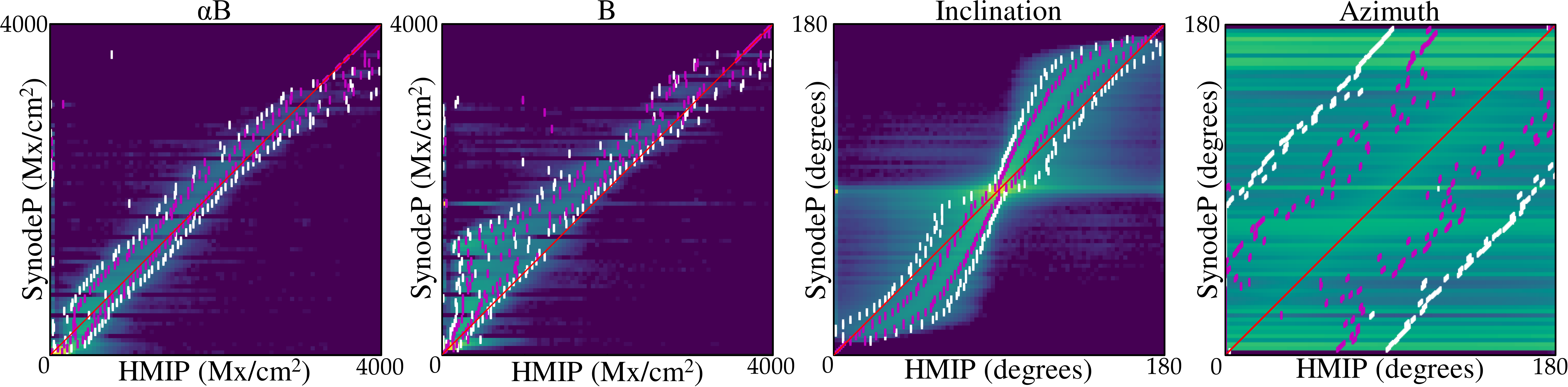}
\caption{Visualizing performance on \textbf{HMI 2016-all}, comparing \synthp against \hmip with {\bf log count} bi-variate histograms. The line $y=x$ depicts perfect predictions, while pink/white dashes show bounds on the left and right (lower and upper) for empirical 67\%/95\% prediction CIs respectively.}
\label{fig:hmi_histograms}
\end{figure*}

The next target, {\it Inclination}/$\gamma$ is also modeled well. In comparison to \hmip, \synthp reproduces the substantially thicker networks of deviations from $90^\circ$ and also the generally more extreme outputs from the \hinode inversions. The distinctions are best seen in difference maps: for instance, in the weaker-field parts of plage, both pipelines infer angles closer to $90^\circ$, but \hmip's is substantially closer. The difference is easier to see in difference maps compared to side-by-side figures. It can also be seen in the bivariate histograms (Figure~\ref{fig:hmi_histograms}), where \hmip's 67\% confidence interval (for values away from $90^\circ$) often barely includes the $y=x$ lines. Finally, \synthp's on-average closer reproduction results in an over lower MAE compared to \hmip. 
Just as it does with $\alpha$, \synthp smooths out small isolated bits of non-$90^\circ$ regions. 

The last target, {\it Azimuth}/$\psi$ generally agrees in regions with good polarization signal, where it reproduces structures well. Variability in placement of visible lines of azimuthal prediction might suggest uncertainty in the exact polarization inversion lines for the model. \synthp's results in weak polarization areas are qualitatively different from both inversions and the network often continues structures observed in stronger polarization into weaker polarization. In these weak polarization regimes, the deep network's predicted distribution over values
ought to be roughly uniform; the emergence of some structure suggests that may be picking up on some signal that reduces the uncertainty slightly.
When tested on larger cutouts containing even weaker polarization regimes, the network often reduces to uniform guesses (seen in Figure~\ref{fig:hmi_histograms}).

\begin{deluxetable*}{lc@{~~~}ccccccccc}[t]
\caption{Results on the \textbf{HMI 2016-All} dataset, comparing \synthp against the \hmip run on full-disk images. We evaluate using MAE and a measure of pixels within a threshold of the target. We adopt the pixel populations presented in H21 where the \textbf{HMI 2016-all} dataset was introduced: ({\bf Disk}) On-Disk, ({\bf Plage}) Plage Pixels, ({\bf AR}) Active Region Pixels, and ({\bf 100+}) Pixels with at least 100 \gauss in the absolute value line of sight magnetic flux density.}
\label{tab:vfisv_results}
\tablewidth{0pt}
\scriptsize
\tablehead{\colhead{Target}&\colhead{Range}&\multicolumn4c{MAE}&\colhead{$t$}&\multicolumn4c{\% Within $t$}\\
&&\colhead{Disk} &\colhead{Plage}&\colhead{AR}&\colhead{100+}&&\colhead{Disk}&\colhead{Plage}&\colhead{AR}&\colhead{100+}}
\startdata
Fill Factor $\times$ Field Strength ($\alpha B$) &[0,$5\!\times\!10^3$] \gauss & 73.2 & 110.4 & 139.0 & 64.7 & 47& 25.3\% & 22.1\% & 20.8\% & 60.6\% \\
Inclination ($\gamma$) &[0,180] \degree &2.4 & 13.9 & 5.4 & 53.1 & 5& 90.1\% & 11.2\% & 53.5\% & 6.1\%  \\
Azimuth ($\psi$) &[0,180] \degree & 41.3 & 15.2 & 9.7 & 55.6 & 7 & 9.7\% & 57.5\% & 75.6\% & 25.4\% \\
\enddata
\end{deluxetable*}
% --- --- End Synode vs. SDO/HMI VFISV Comparison --- ---

% --- --- Synode vs. SDO/HMI VFISV Comparison --- ---

We finally compare how well \synthp aligns with \hmip within the cutouts of the {\bf Paired SDO-Hinode} dataset. Quantitatively, \synthp has a MAE (when compared to \hinp) of 72.4 \gauss for $\alpha B$, 9.4$^\circ$ for $\gamma$, and 24.0$^\circ$ for $\psi$. For all targets, these errors were smaller than \hmip's errors against \hinp, suggesting \synthp is the best full-disk approximation of \hinp available. However, when comparing \synthp directly to \hmip cutouts, we find that \synthp has a MAE of 83.7 \gauss for $\alpha B$, 7.3$^\circ$ for $\gamma$, and 23.2$^\circ$ for $\psi$. This additional comparison shows that \synthp, surprisingly, is often as close to \hmip as it is to \hinp. Given that the quantitative error between \hinp and \hmip is larger for all targets than is the error for \synthp against either \hinp or \hmip, it suggests that \synthp may lie between the two inversion targets.

At first glance, large errors like the $72.4$ \gauss of $\alphaB$ suggest poor performance compared to a prior method such as H21, which garners an error of $9.7$ \gauss in predicting $\alphaB$. However, these errors are not comparable. The largest prediction errors tend to be in strong-field regions, and these strong-field cutouts comprise most of the \textbf{Paired SDO-Hinode} dataset, unlike the full-disks of the {\bf HMI 2016-All} dataset. Instead, the closest related comparison is to evaluate against active regions, where H21 found a MAE of $108$ \gauss in active regions and $41$ \gauss in boxes around active regions.

Overall, our results show that \synthp comes closer to the corresponding \hinp data than \hmip, has systematic differences from \hmip, and demonstrates that \synthm is learning from the \hinp cutouts rather than simply producing a VFISV-like inversion.

\subsection{Evaluating Full-Disk \synthp Inversions} 
\label{sec:synode_vs_hmi}

Our previous section demonstrated that \synthp output was in better agreement with \hinp compared to the agreement between \hmip and \hinp.
We next illustrate the power of the synthetic data products: \synthp uses only \hmi IQUVs as inputs, letting us produce full-disk \synthp inversions for any \hmi observation. Sample qualitative results are shown in Figure~\ref{fig:qualitative_hmi} and additional full disk results appear in Figure~\ref{fig:fig_full}. 
We compare the resulting \synthp inversions with the \hmip. Two possible sorts of failures are possible when moving to full disks: \synthp could fail to generalize beyond cutouts and not produce inversion-like results, and \synthp could demonstrate no systematic differences (improvement) from \hmip.

We show quantitative results on the {\bf HMI 2016-all} dataset in Table \ref{tab:vfisv_results}. These are best examined with the bivariate log histograms in Figure~\ref{fig:hmi_histograms}. Although the {\bf Paired SDO-Hinode} dataset provides a valuable look at difficult-to-predict active regions on the solar disk, the {\bf HMI 2016-all} dataset contains full-disk HMI data, which includes large swaths of low-signal, quiet-pixel regions. Therefore, we evaluate on masks (subsampled from the full disk) from H21 that capture {\bf Plage}, {\bf Active Regions (AR)}, and pixels with at least 100 \gauss ({\bf 100+}). Metrics are the same as mentioned previously in the section comparing \synthp results to \hinp.

On the full disk, \synthp has generally good agreement with \hmip results in $\alpha B$ and inclination, but performs worse when predicting azimuth. Disagreement in inclination rises sharply in plage regions, where just 11\% of pixels are within $5^\circ$. This counter-intuitive trend was also found in a comparison between \hmip and \hinp by \cite{dalda2017statistical}, who ascribed it to differences in treatment of fill factor, which has its strongest effects in plage regions. The strong disagreement in overall azimuth is misleading, since the full-disk numbers are dominated by results in quiet regions, where the azimuth is not well determined. As seen in Figure~\ref{fig:qualitative_hmi}, \synthp tends to produce a constant value rather than noise. In regions with higher polarization signal, such as plage and active regions, azimuth predictions have far better agreement. While average errors show rough agreement, the histograms indicate systematic discrepancies: for instance, as described in Section~\ref{sec:synode_vs_hinode}, \synthm produces inclinations that are substantially further from $90^\circ$. 

In total, this suggests that \synthm is producing values that have rough agreement with VFSIV, but continue to have systematic differences, often in line with \hinp. 

% --- --- Ablations and Comparisons --- ---
\subsection{Ablations and Comparisons} \label{comparison with alternate approaches}
\label{sec:experiments_ablations}

Our cross-satellite observation to inversion prediction task is new and admits a number of alternate solutions. We next quantitatively examine some of these approaches. In this context, ablations are versions of \synthm with pieces removed or substituted. 

\begin{deluxetable}{lc@{~~~}ccccc}[t]
\caption{Variations of \synthm ablated on the {\bf Paired SDO-Hinode} dataset, evaluated using MAE. Regression performs similarly to our proposed classification system, suggesting both methods are reasonable approaches to synthetic inversions. We choose classification as our proposed method as it also easily enables uncertainty quantification. Adding metadata only improves results marginally.}
\label{tab:ablation}
\tablehead{\colhead{Model}&\multicolumn4c{MAE}\\
&\colhead{$\alpha B$ (\gauss)}&\multicolumn1c{$\gamma$ ($^\circ$)}&\colhead{$\psi$ ($^\circ$)}&\multicolumn1c{$\alpha$}}
%\decimalcolnumbers
\startdata
Proposed & 76.1 & 9.4 & 24.0 & 0.12
\\
Regression & 77.7 & 9.3 & 28.9 & 0.11
\\
+Metadata & 74.3 & 9.4 & 23.9 & 0.12
\\
\enddata
\end{deluxetable}

% --- --- AIA --- ---
\begin{figure*}[t]
\centering
\includegraphics[width=\linewidth]{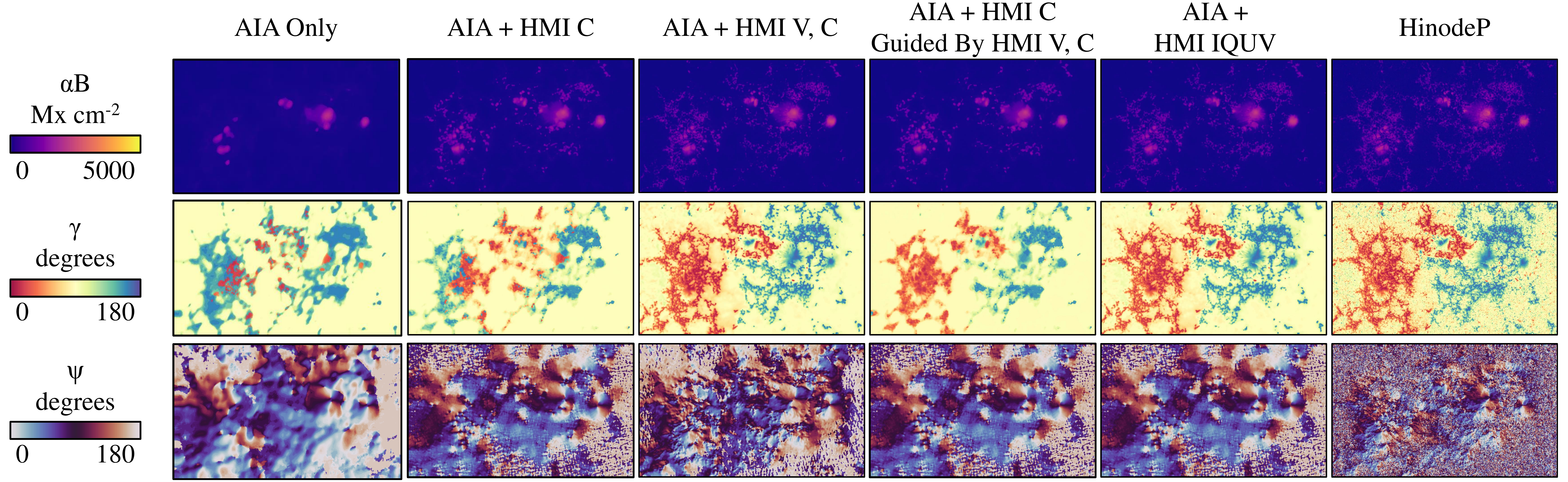}
\caption{Qualitative comparison of  area-averaged magnetic field strength $\alphaB$, inclination $\gamma$, and azimuth $\psi$ on 2014 August 16 15:15:06 TAI from the \textbf{Paired SDO-Hinode} dataset. We show results from varying implementations of \synthm that use different subsets of both \aia and \hmi inputs, with the \hinode inversions on the far right for comparison. Proceeding from columns left to right, we generally see increasing fidelity to the far right column (\hinp) as the amount of polarizaton used as input increases. Of note is the guided AIA+HMI C only inclination output (fourth column), which is far superior to the second column and only relies on an additional bit of information from the third column ($\alphaB$ and $\psi$ outputs remain unchanged from column 2). Guided outputs are a promising means to
create data products from multiple inputs (e.g., AIA+Continuum information plus a single bit from V).
}
\label{fig:varying_inputs}
\end{figure*}

{\it Regression vs Regression-by-Classification.} 
Our core approach is based on regression-by-classification like H21 but several alternatives are viable. As Table~\ref{tab:ablation} suggests, switching to a regression loss produces similar results. Each approach has its own qualitative quirks: classification outputs have minor banding effects.  On the other hand, regression outputs tend to bias towards their means: \synthm with regression losses consistently produces $90^\circ$ inclination and azimuthal predictions across the quiet regions. Classification's binning offers other benefits: it enables confidences, and as we describe in Section~\ref{sec:aia_for_prediction}, enables a post-hoc update of outputs. 

{\it Metadata.}
Training a neural network that can generalize from small image patches recorded by \hinode to full-disk \hmi presents different challenges than H21. We experiment with variants that contain metadata (latitude/longitude), which H21 found to help performance. In practice, we find only a small gain, as seen in Table~\ref{tab:ablation}. While experimenting with regression, we found that metadata-trained networks occasionally mis-predict in the polar regions where the training data include significantly sparser \hinode data that were successfully aligned.  Hence, in the interest of avoiding other metadata issues, we do not include metadata.

{\it Independently-Determined Fill Factor.} 
As described in Section~\ref{sec:targets}, the area-averaged field strength $\alpha B$ is the product of the fill factor $\alpha$ and intrinsic field strength $B$ and is the field strength quantity produced by the \hmip. In our experiments, we predicted this with a network dedicated to predicting $\alpha B$. We examined how this compares with independently modeling $\alpha$ and $B$. We find independent prediction of $\alpha$ and $B$ to perform similarly to prediction of the joint quantity $\alphaB$, achieving a MAE of $77.6$ \gauss compared to $76.1$ \gauss from the joint model.
% --- --- End Comparison With Alternate Approaches --- ---

\begin{deluxetable*}{lc@{~~~}ccccc}
\caption{The \aia imager on {\it SDO} co-observes with \hmi, and \synthm lets us  explore how varying combinations of these inputs impact synthetic inversion performance and thus what information can be extracted from each input. Results are reported for the \textbf{Paired SDO-Hinode} dataset.
Group one are reference results. Group two lacks polarization; different subsets can achieve good performance on $\alphaB$ but do poorly on $\gamma$ and $\psi$. Group three varies polarization inputs. Just $V$ can do well on all but $\psi$, but best results are obtained with Q and U included. The final group reports results on $\gamma$ by a model without polarization data when its direction is picked by model with polarization data; this guided model shows large improvement.
} 
\label{tab:aia}
\tablewidth{0pt}
\scriptsize
\tablehead{\colhead{Inputs}&\multicolumn4c{MAE}\\
&\colhead{$\alpha B$ \gauss}&\multicolumn1c{Inclination $\gamma$ (${}^\circ$)}&\colhead{Azimuth $\psi$  (${}^\circ$)}&\multicolumn1c{ $\alpha$}}
\startdata
\hinp mean value in training data  & 342.7 & 26.8 & 42.2 & 0.21 \\
\synthp i.e., \synthm-(HMI Continuum + IQUV) & 76.1 & 9.4 & 24.0 & 0.12 &\\
\hmip & 109.7 & 12.5 & 28.3 & 0.49 \\
\hline
\synthm-(AIA) & 171.0 & 25.5 & 31.7 & 0.14 \\ 
\synthm-(AIA STEREO Subset: 171\AA, 193\AA, 304\AA) & 235.5 & 28.5 & 33.9 & 0.18 \\
\synthm-(AIA UV Subset: 1600\AA, 1700\AA) & 149.6 & 28.2 & 32.5 & 0.13 \\
\synthm-(AIA + HMI Cont.) & 107.6 & 28.7 & 37.5 & 0.13 \\ \hline
\synthm-(AIA + HMI V) & 80.0 & 10.1 & 31.1 & 0.12 \\ 
\synthm-(AIA + HMI Cont.,  V) & 77.9 & 10.0 & 32.1 & 0.12 \\
\synthm-(AIA + HMI Cont., IQUV) & 74.6 & 9.4 & 24.0 & 0.12 \\
\hline
\synthm-(AIA + HMI Cont.) Guided By  AIA  + HMI Cont, V & -- & 16.5 & -- & --
\enddata
\end{deluxetable*}

\subsection{Using AIA for Prediction}
\label{sec:aia_for_prediction}

We next explore how to use data from other instruments as input for \synthm with the continued goal of predicting \hinp. Specifically, we examine combining various subsets of \aia and \hmi data to serve as input. We explore \aia by itself (and subsets thereof), and then with increasing number of Stokes vector components added, starting with circular polarization (V) and continuum intensity. Continuum intensity is calculated from I since HMI does not sample the true continuum~\citep{couvidat2012line}. We focus on V since it has a higher signal-to-noise ratio (SNR) than [Q,U].  One might imagine a hybrid sensor that uses high-SNR components of various instruments to produce reduced noise/strong confidence predictions. In each case where we use a particular Stokes component, we include all six \hmi pass bands. We report quantitative results in Table~\ref{tab:aia}. At the top in Group 1, we report a few numbers to give context: the mean of the \hinp output; the proposed variant of \synthm, which uses Continuum and all Stokes components; and \hmip. We show qualitative results in Figure~\ref{fig:varying_inputs}.

The second group in Table~\ref{tab:aia} shows
\synthm variants without access to polarization information and which use only \aia channels or continuum. We test four variants. The first ({\it AIA}) is the set of 9 \aia channels (all channels other than 4500\AA~white light continuum). The second ({\it AIA STEREO Subset}) includes the 3-channel EUV subset (191\AA, 193\AA, 304\AA) that matches observations from {\it STEREO}/SECCHI/EUVI. This tests what information can be extracted from information currently obtainable on the far side of the Sun. The third ({\it AIA UV Subset}) are the two UV channels measured by AIA (1600\AA, 1700\AA). These originate in the chromosphere, which is substantially closer to photosphere. Finally, we add \hmi Continuum information, which provides observations of the photosphere, but no polarization signal.

\begin{figure}[t]
\centering
{
\includegraphics[height=1.65in]{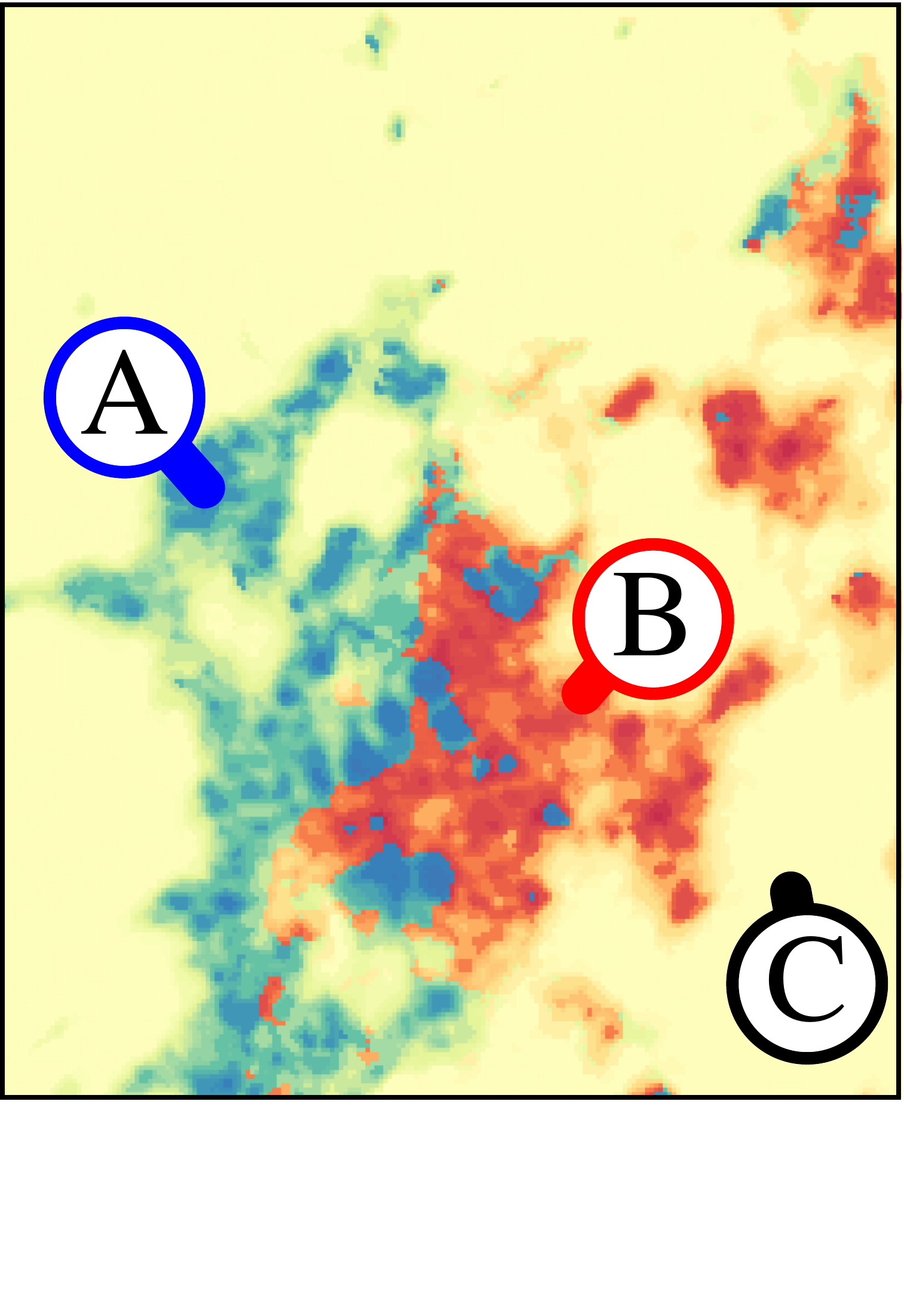} 
\includegraphics[height=1.65in]{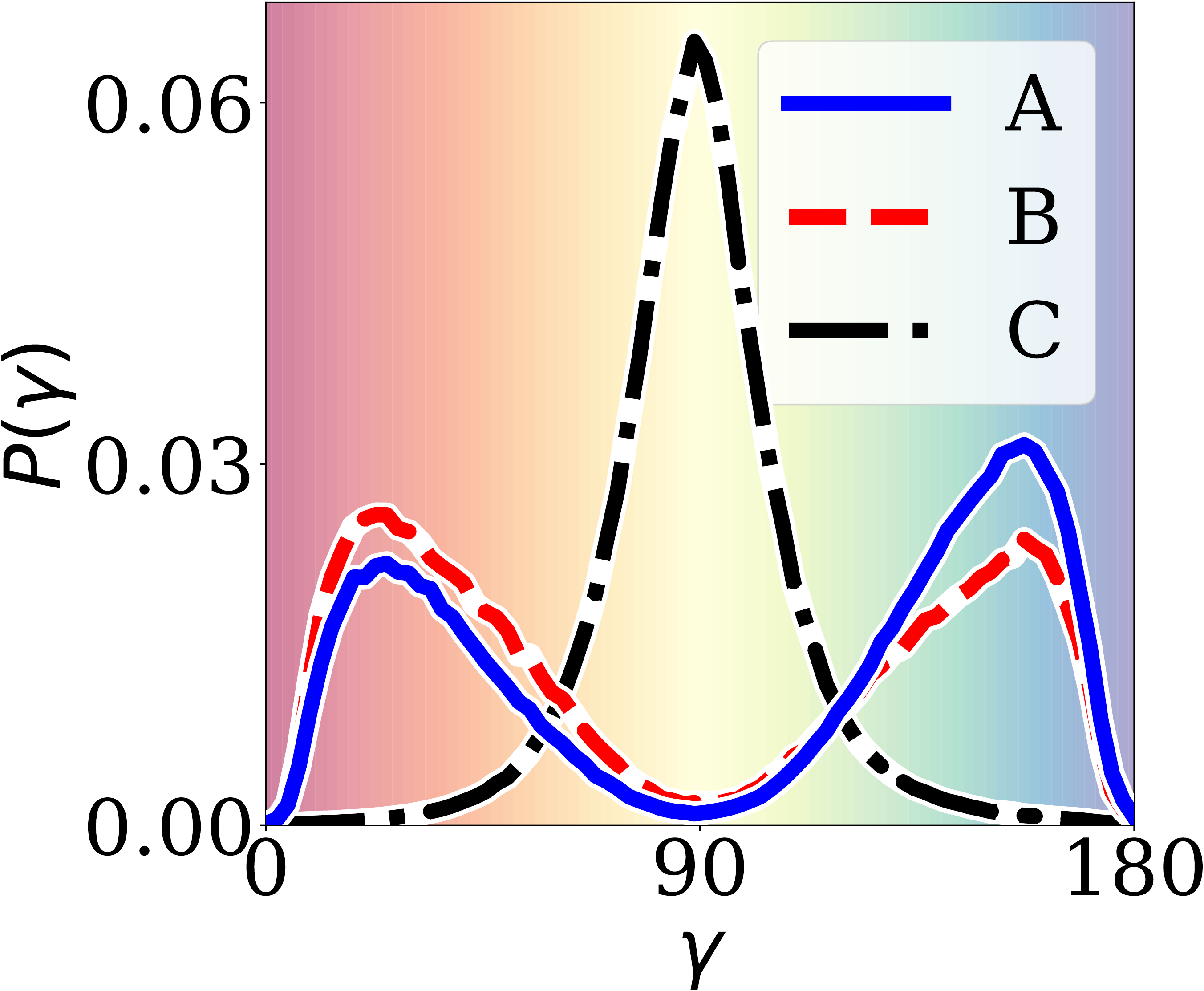}
}
\caption{Sample distributions over $\gamma$ predicted by a \synthm network alternative using \aia and Continuum inputs on 2014 August 16 15:15:06TAI, like Fig.~\ref{fig:varying_inputs}. In regions with non-$90^\circ$ $\gamma$ (A,B), the distribution becomes bimodal. In regions where $\gamma$ is near $90^\circ$ (C), the distribution is unimodal. However, given one bit of information (e.g., from V) about direction, one can select the correct direction and mode in A,B.
}
\label{fig:bimodal}
\end{figure}

While the polarization-free variants of \synthm can estimate quantities related to field strength (e.g., $\alpha B$ and $\alpha$), they are unable to provide detailed information about magnetic field direction, $\gamma$ and $\psi$. This is consistent with underlying physics: flipping the field polarity would produce identical intensity data. As Fig.~\ref{fig:bimodal} illustrates, in regions where $\gamma$ is likely far from $90^\circ$ (e.g., plage, active regions) the network's belief splits into two modes: one below $90^\circ$ and the other above. Without additional signal, the selected mode is more or less random. While our results seem to contradict ones found by~\cite{Jeong2020} (where the sign of the LOS magnetogram could be predicted from STEREO-like data), we suspect both model and data are at play. \cite{Jeong2020} trained a conditional GAN model~\citep{pix2pix2017}, which prefers predictions a second deep network thinks appear globally similar (in addition to per-pixel losses). Our mid-structure sign-flips are unlikely and would be eliminated with a GAN loss. Second, our test data is ${\ge}68$ days {\it after} the training sample, while the test data used by ~\citet{Jeong2020} includes randomly sampled months within the training period, providing the tests with training data on both sides of the month and potentially enabling a better guess of the sign.

\begin{figure*}[t]
\centering
\begin{tabular}{cc}
\includegraphics[width=0.48\linewidth]{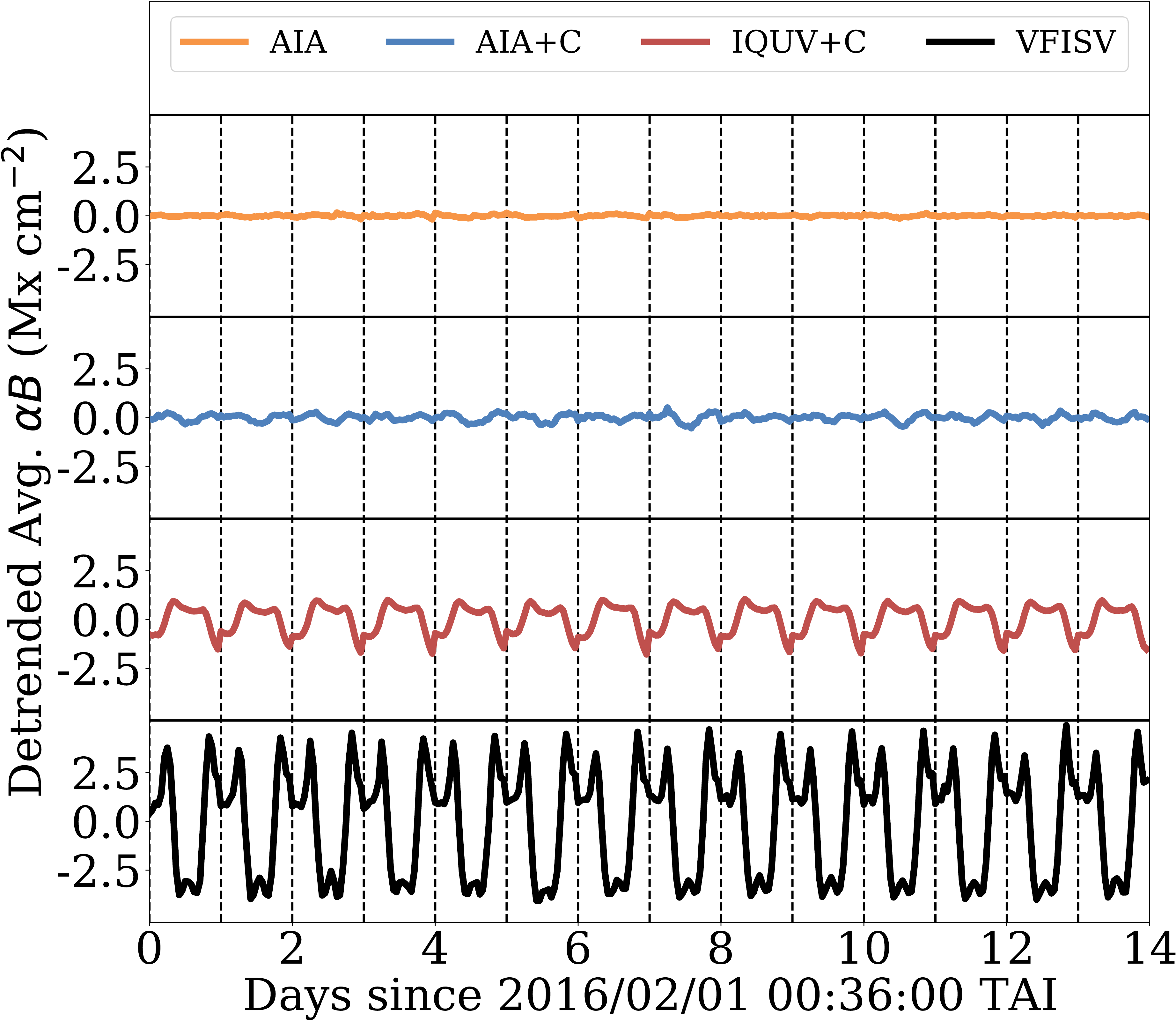} &
\includegraphics[width=0.48\linewidth]{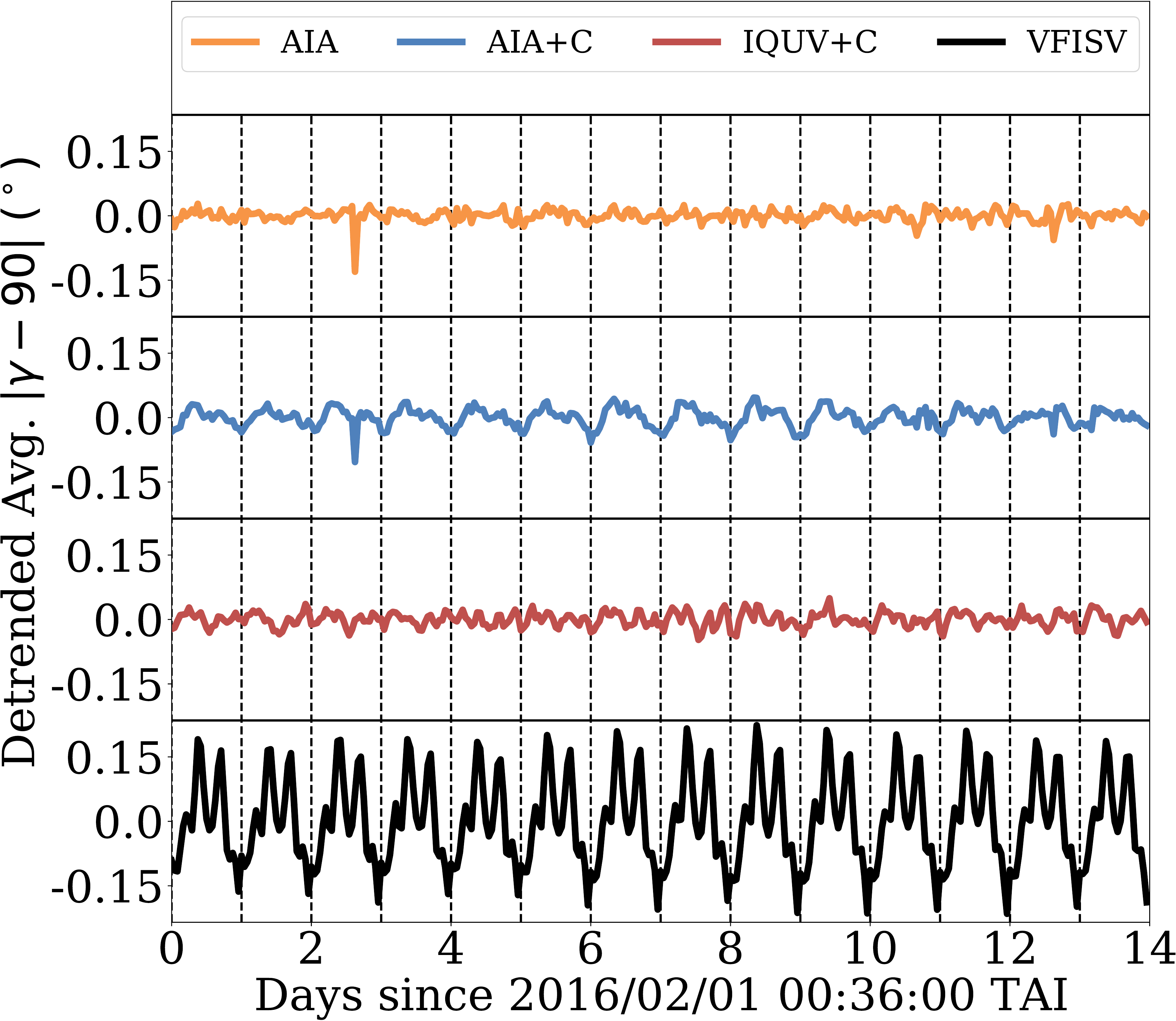} \\
\end{tabular}
\caption{The pixel-averaged field strength (left) and deviation from inclination=$90^\circ$ (right) are followed across the disk for each inversion, plotted as a function of time.  \hmip is known to have a 24-hour oscillation in these quantities. We compute the average on the disk, removing the limb. We linearly detrend the data each day to better show the oscillations and bring $\alphaB$ from \synthp and \hmip VFISV to the same values. \synthm-based models have smaller oscillations, and \synthm with AIA has close to no oscillation. A synthetic inversion with minimal oscillations 
may be of use in tasks that need minimal oscillations (at the price of reduced fidelity). 
Moreover, per-pixel difference maps across the full disk comparing inversions against synthetic inversions, when viewed over days and weeks, may aid in elucidating where and to what degree oscillations on the disk occur.}
\label{fig:fig_iquvs_bring_oscillation}
\end{figure*} 

We also observe that the data are substantially more blurry than using the \hmi Stokes observations. We hypothesize that the blurriness happens because the EUV observations originate at multiple heights above the photosphere and so are intrinsically misaligned. Results from models that use only AIA are substantially sharpened by adding continuum observations that originate in the photosphere.

Inclination angle can be obtained by directly adding polarization information as an input and \synthm lets us quantify the performance improvement, shown in group three of Table~\ref{tab:aia}. Adding V to the polarization-free models dramatically improves estimates of $\gamma$, halving the error rate regardless of whether one has access to continuum data. We note though that adding \aia data to models that have access to the full Stokes vector (e.g., \hmi IQUVs and Continuum) does not meaningfully improve results. This is intuitive, since most of the AIA channels do not originate in the photosphere. 

The bimodal nature of \synthm's predictions for $\gamma$ without polarization information also suggests that one could use an additional cue to {\it guide} selection of one of the predicted modes. For instance, in Fig.~\ref{fig:bimodal}, if one knew whether the angle was above-or-below $90^\circ$ (i.e., only the polarity of the line-of-sight magnetic field), one could select the correct angle. In practice, we solve this by finding the side with a system that has access to polarization, 
setting probabilities in the non-agreeing parts of the output space to $0$, and re-normalizing. 

We provide a proof of concept for inclination using \synthm with \aia and \hmi Continuum as the base system, and guide it with the predictions from a model that also uses V. Specifically, we force a prediction from the half predicted by the more capable system -- note that only a single bit of information from $V$ is used. The resulting system's inclination substantially improves, as seen in Figure~\ref{fig:varying_inputs}. The experiment suggests that one could construct virtual observatories, with different instruments providing different pieces of information but each less substantive than required for a full vector spectropolarimeter. 

\subsection{Model Results Across a Time-Sequence}
\label{sec:network_behavior_across_time}

\hmip outputs are known to have non-physical oscillations in many quantities~\citep{Hoeksema2014,schuck2016achieving}, and the base system we use was shown by H21 to reproduce these oscillations. Oscillations manifest in multiple ways, including the on-disk average magnetic flux density $\frac{1}{n}(\sum_p (\alpha B)_p)$ and inclination deviation from $90^\circ$, $\frac{1}{n}(\sum_p |\gamma_p-90|)$. These quantities are known to vary with a 24 hour period (plus harmonics). The \hinode MERLIN inversion pipeline, on the other hand, does not exhibit these oscillations. We stress that while we show plots across time, each observation is produced independently.

Figure~\ref{fig:fig_iquvs_bring_oscillation} shows averaged $\alpha B$ and inclination ($\gamma$) deviation from $90^\circ$ over a two-week period for \hmip and multiple \synthm variants. All time series are linearly detrended day by day (i.e., by subtracting a linear daily fit) to make analysis easier: \hmip has a substantially higher average field strength in quiet regions than both \synthp and \hinp and the regularity is more difficult to see when mixed with natural evolution of the field.  We report the 95\% data range to quantify the range of the oscillation and test for its presence by computing the Durbin-Watson \citep[DW;][]{durbin1950testing} autocorrelation statistic with a 24-hour lag. The DW statistic of a signal is a normalized sum of squared differences between a signal and its lagged copy, $\sum_{i=1}^{N-24} (S_i - S_{i+24})^2 / \sum_{i=1}^{N-24} S_i^2$. Values near $0$ indicate strong autocorrelation, and values near $2$ indicate no autocorrelation.

\hmip has, by far, the strongest oscillations, with DW statistics $<0.01$ and a large 95\% data range of $8.3$ \gauss ($\alpha B$) and $0.38^\circ$ ($\gamma$).
\synthp, or training \synthm on \hinode outputs and \hmi inputs (IQUV + Continuum intensity) still yields oscillations (DW $0.02$ for $\alpha B$ and $0.72$ for $\gamma$), but the magnitude is substantially dampened (95\% range: $2.4$ \gauss, $0.07^\circ$). The continued presence of oscillations in \synthp output when trained on \hmi input data suggests that the source of the oscillations is not incorrect modeling in the \hmi VFISV inversion pipeline itself, {\it i.e.} the Milne-Eddington assumptions per se. Instead the re-emergence of the oscillations suggests that they  originate in a loss of information in the measurements that leads to different times of day or different orbital velocities introducing characteristics that cannot be corrected for by the LOS velocity fit in the inversion procedure. 
When only \aia data are used in \synthm, the variations disappear (DW statistics $2.1$ for $\alpha B$ and $1.8$ for $\gamma$). This result is expected since neither the input, \aia, nor the output, \hinp, are known to have the temporal oscillations.
On the other hand, if Continuum intensity is added, the oscillations re-emerge (DW: $0.7$ for $\alphaB$ and $0.573$ for $\gamma$), albeit with a much diminished range for $\alphaB$ (95\% range: $0.6$ \gauss, $0.8^\circ$). Excitingly, \synthm using AIA+C has about as good agreement with \hinode on $\alphaB$ in comparison to \hmip, but has substantially reduced oscillations. This reduction in oscillation comes at the price of poor prediction of inclination and azimuth due to a lack of polarization information.

In all cases, \synthm models have diminished oscillations, suggesting that this system may serve as a mechanism for obtaining oscillation-free estimates of the magnetic field. Moreover, Figure~\ref{fig:fig_iquvs_bring_oscillation} and Table~\ref{tab:aia} show that \synthm with varying inputs (e.g., just using continuum) can form a trade-off between fidelity to \hinode and the presence of oscillations.
% --- --- End Model Results Time Sequence --- ---

\subsection{Limitations}
\label{sec:limitations}

\begin{figure}
    \centering
    \includegraphics[width=\linewidth]{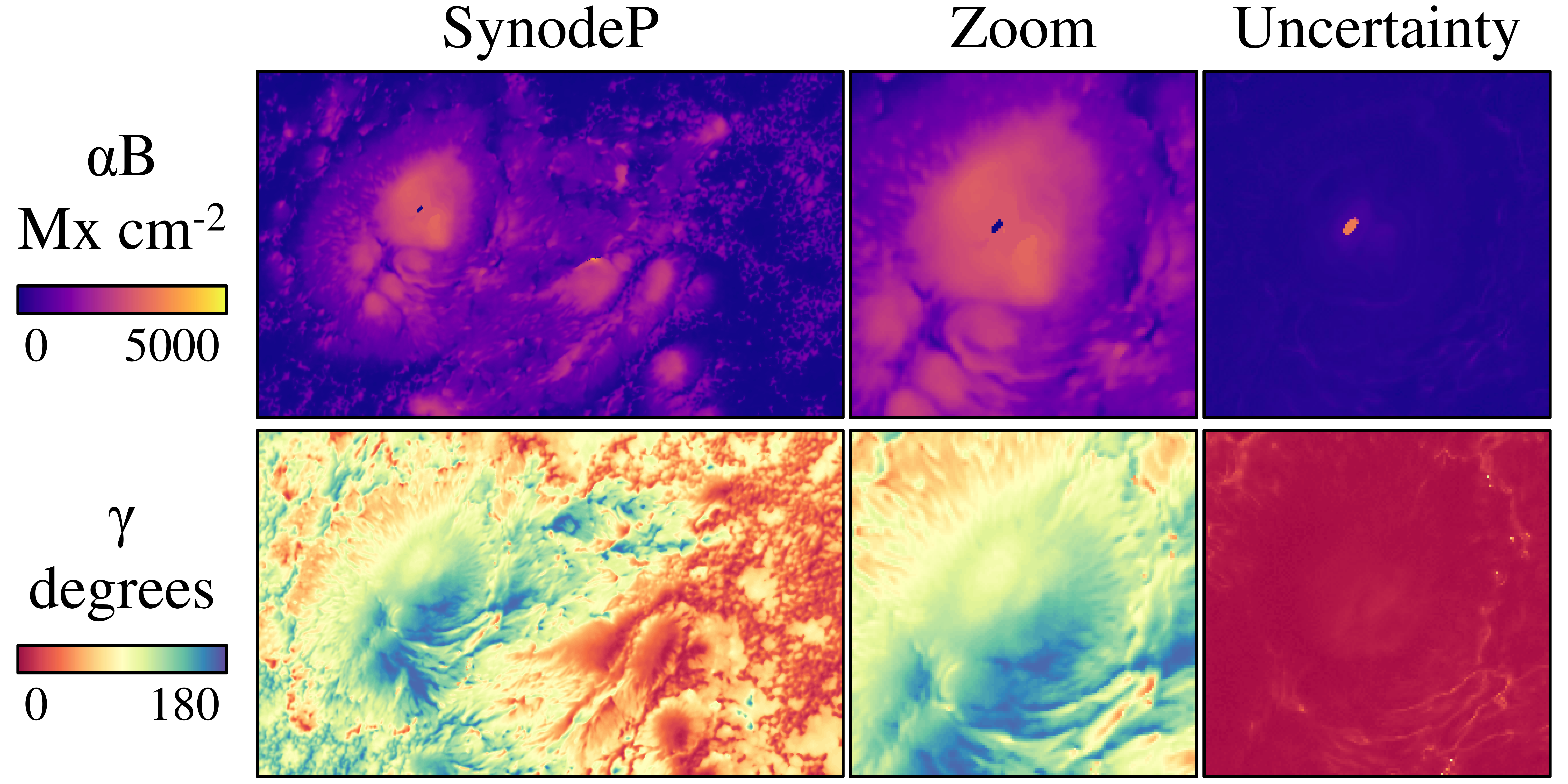}
    \caption{\synthp failure modes (2014 October 21 13:01:27 TAI): like H21, the proposed system can have trouble in the center of large sunspots, especially in $\alphaB$. In this case, the network is uncertain about its prediction. Left: $250.5\arcsec \times 147.5\arcsec$; right: $75\arcsec \times 75\arcsec$ zoom-in on prediction and uncertainty). These holes also exist in some outputs of \hmip and \hinp, and show the \synthp outputs are akin for both good and bad performance.}
    \label{fig:sunspotfail}
\end{figure}

We conclude our experiments by discussing some limitations of the \synthm approach. Some are due to the cross-instrument alignment and some are inherited from H21, e.g., the use of multiple networks, need for post-inversion disambiguation, and saturation of field, albeit at a higher value, etc. We describe particularly important ones below:

{\it Large Sunspots.} The proposed network can struggle with the centers of large sunspots, as seen in Fig.~\ref{fig:sunspotfail} and paradoxically produce low field strength. A more involved discussion is given by H21, but strong Zeeman splitting and low intensity leads to little signal near line center. While weak signal can be explained by high field strength, it can also be explained by the far more frequent case of a lack of magnetic field. \synthm identifies that it is far more uncertain about this prediction compared to the rest of the prediction. Nonetheless, we note that smaller pixel-wise inversion failures are present in the outputs of both \hmip and \hinp in a later observation of the same active region as shown in Figure~\ref{fig:hinode_synthm_vfisv}. This phenomena is also reported in~\cite{Hoeksema2014}.

{\it Alignment between \hinode and SDO.} \synthm depends on alignment between input and output data products. Our results suggests that alignment is generally good, but the results have a slight blurriness, and small few-pixel structures in outputs like fill factor and inclination are erased. The erasure of small features suggests remaining misalignment: if similar pixels are often misaligned at training time, the loss-minimizing behavior puts more weight on more common outputs.
In classification, the network can split its belief into multiple bins; during inference time, the mode (rather than less-likely prediction) is selected. The \synthm with \aia inputs has more issues with alignment off disk-center due to varying heights at which the observations originate. We leave alignment at varying altitudes, and other improvements, to future work.

{\it Integrated Evaluation with Downstream Applications.} While \synthm outputs do appear to exhibit weaker 24-hour oscillations, it is possible that \synthm introduces other biases and artifacts. Further investigation is needed to identify these, including integration of outputs from \synthm into potential downstream applications, such as done in \cite{Jeong2020}.

%% file: sec_conclusions.tex
We present \synthm (Synthetic Inversion Algorithm), a deep learning-based approximation for Stokes vector inversion. We have applied it to build a model that can map from \hmis observables to \hinode pipeline outputs (\hinp), yielding a data product we name \synthp.
Our results in Section~\ref{sec:synode_vs_hinode} show that the \synthp system is able to produce outputs that resemble \hinp more than the \hmi pipeline results (\hmip) resemble \hinp on cutouts typically centered on active regions. When inferred on the full-disk, \synthm produces results that broadly agree with \hmip, but exhibit systematic differences (Section~\ref{sec:synode_vs_hmi}) that often resemble patterns in differences found between \hinp and \hmip. While full-disk \hinp cannot be obtained currently, the results suggest that \synthm may behave similar to what \hinp would find if it scanned the full disk. Our results in Section~\ref{sec:experiments_ablations} suggest that multiple variants of our method are effective but have different trade-offs. Our results open the doors to a wide variety of applications using \synthm-like systems. For instance, as shown in Section~\ref{sec:aia_for_prediction}, our approach can generalize to use \aia. As our guided decoding demonstrates, multiple sources of information can be used to construct inversions. Similarly, \synthm results, even with \hmi inputs, exhibit far weaker 24-hour oscillation artifacts (as seen in Section~\ref{sec:network_behavior_across_time}), and may offer a path forward to oscillation-free full-disk inversions.

Our investigation is part of a growing field of work using deep learning to predict pixel-wise solar data. This has included (among others) work that predicts between UV/EUV observations \citep{Salvatelli2019}, from magnetograms to UV/EUV observations \citep{Galvez2019,Park2019}, from UV/EUV observations to magnetograms \citep{kim2019solar,Jeong2020}, magnetograms from observed spectra \citep{higgins2021fast,liu2020inferring,socas2001fast}, and magnetograms from synthesized spectra \citep{Ramos19}. In terms of high-level motivation, \synthm differs from all of these by mixing instruments on different satellites during training.

The experimental methodology, however, underscores a difference in perspective from many of these works. Many previous approaches have focused on predicting lower-resolution (e.g., $512^2$) data products and sometimes with values that have been clamped to a fixed range. These experiments are important first steps. \synthm aims to produce inversions far upstream and as close as possible to level 1 observables, and at native instrument resolution ($4096^2$) with effectively no min/max value clamping. Current global MHD models like AWSoM~\citep{van2014alfven} cannot benefit from this high resolution due to their uses of grid sizes on the order of $1^\circ$; however future ones may benefit, and local MHD and NLFFF models can immediately benefit from the increased cadence and range of targets available for investigation. Moreover, the analysis of upstream data lets us analyze known artifacts like oscillation at their source. We hope that the community will jointly explore the challenges and opportunities of predicting and evaluating at native resolution and upstream.

From a perspective of Stokes inversion, we have demonstrated that \synthm can produce output from \hmi spectra that more closely resemble \hinode MERLIN inversions (i.e., \hinp) than do \hmi VFISV inversions (i.e., \hmip). If \hinp is believed to have better agreement with physical reality, then this agreement suggests that \synthm may provide a path towards obtaining results that are closer to reality than \hmip (in the specific sense of MAE). We stress though, that MAE is not the only criterion. For instance, the community has a ten year head start on understanding the idiosyncrasies and limitations of \hmip compared to those of synthesized inversions like \synthp.  This paper identifies some of \synthp's idiosyncrasies, such as the lack of a noisy pattern in azimuth in weak-polarization areas. However, others are yet to be discovered and will likely require further tests with downstream applications. That said, \synthp's better per-pixel agreement, full-disk separation of $\alpha$ and $B$, and reduced oscillation suggests that tremendous progress has been made, and that integration with downstream applications may be fruitful.

\synthm's ability to produce \hinp-like inversions suggests that there is additional information in the \hmi spectra that is not being extracted in the current HMI pipeline. The information may be in the per-pixel spectra, but pipeline VFISV may be unable to find it, either because the $\chi^2$ agreement with \hmi spectra does not provide a suitable gradient for ME-inversion minimization or due to pipeline variable decisions (e.g., setting fill factor to 1). An alternate hypothesis is that additional information is in the joint spectra of a set of pixels (and indeed, anecdotally, we have found that networks that depend on larger regions of the image perform better). One advantage of learning-based Stokes inversion techniques is that they can easily leverage information like the surrounding pixels without needing a physical model. A full treatment is beyond the scope of this paper, but our initial results suggest that \synthm offers data-driven tools for peering into what information can and should be extracted from which data.

\synthm's inversion of the \hmi spectra provides an opportunity. Since the system requires only \hmi spectra, it enables the full-disk \hinp-like inversion of all of \hmi's historical Stokes vectors, which in turn presents many interesting opportunities. First, one can systematically explore areas where \hmip and \synthm disagree, illustrating areas that may have been incorrectly inferred (while being unable to distinguish which is wrong). Our initial results certainly demonstrate differences in terms of interpretation of nearly every region, but a systematic and detailed study is now possible (although beyond the scope of this paper). Second, one can obtain \hinp-quality data for multiple parts of the disk, which enables many interesting applications. Since \hinode has a limited field of view, it is difficult to capture, for example, the very beginning of AR emergence. Moreover, since \synthm requires training, one could train a \synthm using higher cadence Stokes vectors obtained by \hmi, up to 90s. 

Finally, the \synthm results point to a number of future opportunities. At present, the ability to ingest data from multiple instruments and the ability to reproduce many features of the targeted inversion result suggests that one may be able to use various observables from multiple instruments operating at different times to create a consistent, high-quality, long-term record of the solar magnetic field.
Looking toward the future, the ability to produce full-disk \hinode-like inversions from smaller regions suggests that future missions may be able to provide best-of-both-worlds systems by combining data from instruments having higher spatial and spectral resolution with complementary instruments having larger fields of view, faster cadence, and greater continuity. Our guided decoding of inclination from \aia and continuum, where a single bit (toward-vs-away) comes from Stokes {\it V}, suggests that one may be able to have multiple simpler instruments contribute various input components, which may enable innovative approaches to obtaining improved inversion results. There is, of course, work to be done, but \synthm presents strong first steps.

%% file: sec_acknowledgements.tex
This work is an interdisciplinary collaboration run by the NASA Heliophysics DRIVE Science Center (SOLSTICE) at the University of Michigan under grant NASA 80NSSC20K0600. It is also supported by the Michigan Institute for Data Science, which has awarded authors with a Propelling Original Data Science grant. GB and KDL of NWRA also acknowledge NASA/GSFC grant 80NSSC19K0317.

All SDO data used are publicly available from the Joint Science Operations Center (JSOC) at Stanford University supported by NASA Contract NAS5-02139 (HMI), see \url{http://jsoc.stanford.edu/}.
Hinode is a Japanese mission developed and launched by ISAS/JAXA, with NAOJ as domestic partner and NASA and STFC (UK) as international partners. It is operated by these agencies in co-operation with ESA and NSC (Norway). Data and models used will be archived at the U-M Library Deep Blue data repository. All datasets will be given Digital Object Identifiers (DOIs).